\documentclass[12pt]{amsart}
\theoremstyle{plain}
\numberwithin{equation}{section}
\newtheorem{Theorem}{Theorem}
\newtheorem{Lemma}[Theorem]{Lemma}

\newtheorem{Proposition}[Theorem]{Proposition}

\newtheorem{Remark}[Theorem]{Remark}

\begin{document}

\title[Trace formula]
{Trace formula and spectral Riemann surfaces for a class
of tri-diagonal matrices}

\author{Plamen Djakov}

\author{Boris Mityagin}

\pagestyle{myheadings}
\markboth{}{}

\address{Department of Mathematics,
Sofia University,
1164 Sofia, Bulgaria}

 \email{djakov@fmi.uni-sofia.bg}

\address{Department of Mathematics,
The Ohio State University,
 231 West 18th Ave,
Columbus, OH 43210, USA}
\email{mityagin.1@osu.edu}

\thanks{The authors thank the Erwin Schr\"odinger
International
Institute for Mathematical
Physics for the hospitality in September 2005.}

\begin{abstract}
For tri-diagonal matrices arising in the simplified Jaynes--Cummings
model, we give an asymptotics of the eigenvalues, prove a trace formula
and show that the Spectral Riemann Surface is irreducible.
\end{abstract}

\maketitle

MSC: 47B36

\section{Introduction}
We consider one-sided tri-diagonal matrices of the form $L+zB,$
where
\begin{equation}
\label{1}
L = \left [
\begin{array}{ccccc}
q_1 &  0 & 0 & 0   &   {\cdot}   \\
0 &  q_2 & 0 &  0  &   {\cdot}  \\
0  & 0 & q_3 & 0 &     {\cdot} \\
0 &  0&   0& q_4 &   \cdot \\
\cdot &\cdot &\cdot &\cdot& \cdot   \\
\end{array}
\right ],
\qquad
B =
\left [
\begin{array}{ccccc}
0 &  b_1 & 0 & 0   &   {\cdot}   \\
c_1 &  0 & b_2 &  0  &   {\cdot}  \\
0   &  c_2 & 0 & b_3 &     {\cdot} \\
0 &  0&   c_3 & 0 &   \cdot \\
\cdot &\cdot &\cdot &\cdot& \cdot   \\
\end{array}
\right ]
\end{equation}
and study their spectra in the case where
the diagonal matrix majorities the off--diagonal one
in the sense of the following condition (or some version of it)
\begin{equation}
\label{2}
|q_k|\to \infty, \quad \frac{(|b_k|+|c_k|)^2}{|q_k q_{k+1}|} \to 0.
\end{equation}

There is a vast literature (see \cite{JN,JNS,DP,S2} and the
bibliography therein)
devoted to a broad range of questions
on these matrices and
the corresponding operators in $\ell^2 (\mathbb{N}).$
We will be concerned with the following three questions.\vspace{3mm}

1. Spectra $Sp(L+zB).$  Of course,
$Sp(L) = \{q_k, \; k=1,2,\ldots \}$ and
$$ Le_k = q_k e_k, \quad k=1,2,\ldots ,
$$
where $\{e_k\}_1^\infty $ is the canonical orthonormal basis in
$\ell^2 (\mathbb{N}).$
Under the condition (\ref{2}) the spectrum
$Sp(L+zB) $ is discrete as well
(see, e.g., Lemma 8 in \cite{DM12}, or \cite{JN01,T2}),
and
$$ Sp(L+zB) = \{E_n (z)\}_1^\infty,
$$
where, for each $n,$
$E_n (z) $ is an analytic function at least for small $|z|,$
i.e., in the disk $|z|< R_n $ for some $R_n >0.$

(1.A) {\em How large could $R_n $ be chosen?}

Let us mention that in the case of Mathieu operator
H. Volkmer \cite{V1} proved that $R_n \asymp n^2 $
(see further discussion in Section 7.1--7.3).

(1.B) {\em What is the asymptotic behavior of $E_n (z) $ if $z$
is bounded, say $|z| \leq R, $
and} $n \to \infty? $
\vspace{3mm}

2. Under the conditions (\ref{2}) and some further assumptions
on the sequences $q, b, c$
one can introduce the regularized trace
$$
tr (z) = \sum_{n=1}^\infty \left (E_n (z) - q_n \right )
$$
as an entire function -- see Definition in Section 5.4.

Can we evaluate it in specific examples?  \vspace{3mm}

3. Spectral Riemann Surface of the pair $(L,B) \in (\ref{1}), (\ref{2})$
is defined as
$$
G= \{ (\lambda, z) \in \mathbb{C}^2 \;: \quad
(L+zB)f = \lambda f,  \quad
 f \in \ell^2 (\mathbb{N}), \; f\neq 0 \}.
 $$

F. W. Sch\"afke proved that in the case of
the Mathieu equation
$$
 -y^{\prime \prime} +z (\cos 2x) y = \lambda y, \quad \text{i.e.,}
 \quad
 L=- (d/dx)^2, \;\; By =  (\cos 2x) y,
$$
the Spectral
Riemann Surface is irreducible
\cite{MSW}, pp. 88--89;
see also \cite{V4}.
We use Sch\"afke's scheme to prove
that the Spectral Riemann Surface  $G$
is irreducible in the case of the simplified Jaynes--Cummings model
(Theorem 3).

We focus our attention on special tri-diagonal matrices which are
motivated by the analysis of second order differential operators
in the framework of Fourier method. \vspace{3mm}

Example 1. Let
\begin{equation}
\label{6.1}
 q_k = k^2, \quad b_k = c_k = k^\alpha, \quad 0 \leq \alpha < 2.
\end{equation}
If
\begin{equation}
\label{6.2}
\alpha = 0,
\end{equation}
we have the Mathieu matrices, and if
\begin{equation}
\label{6.3}
\alpha = 1/2,
\end{equation}
we have the simplified Jaynes--Cumming matrices
that have been considered by A. Boutet-de-Monvel, S. Naboko and
L. Silva \cite{BNS}. \vspace{3mm}

Example 2. More general $q,$
$$ q_k = k^\gamma, \quad b_k = c_k = k^\alpha, \quad \gamma \geq \alpha+ 1/2.
$$
The case $ \gamma = 1, \alpha = 1/2 $ comes from
the Jaynes--Cumming model
(see E. Tur \cite{T1,T3}). \vspace{3mm}

Example 3. The Whittaker--Hill matrices (see \cite{MW}, Ch.7,
and \cite{DM12})
\begin{equation}
\label{7.1}
q_k = k^2 \;\; \text{or}\; \; (2k+1)^2, \quad
b_k = t - k, \; c_k = t+k, \quad t \geq 0 \;\; \text{fixed}.
\end{equation}
We do not provide details about the Fourier method or
the gauge transform which lead us from the differential operator
$$
- y^{\prime \prime} +(a \cos 2x + b \cos 4x )y
$$
to the matrices (\ref{1}) with (\ref{7.1}). See \cite{Ince, UA, MW, DM12}.
In Section 7.1, Propositions \ref{prop17} and \ref{prop18},
we use results about differential operators
\cite{V1,V98,V2}
to find
asymptotics of the radius of analyticity $R_n $ in the case of matrices
(\ref{7.1}).

The matrices (\ref{6.1})--(\ref{6.3})
and (\ref{10.1}), (\ref{10.2}) is
the main object of interest
in this paper. Now we spotlight some of its results.
Below $E_n (z) $ means the $n$-th eigenvalue of
$L+zB.$

\begin{Theorem}
\label{thm1}
Suppose 
(\ref{10.1}) and  (\ref{10.2}) 
with $ 0 \leq \alpha \leq 1/2 $ hold, and
$\lim_k b_k c_k k^{-1} = \ell $ exists
 for $ \alpha =1/2. $ Then, for $ \alpha \in [0, 1/2],$
the regularized trace $tr(\alpha,z)$
is well--defined entire function, and
\begin{equation}
\label{7.2}
tr(\alpha,z) \equiv \sum_1^\infty (E_n (z)- n^2) = \left \{
\begin{array}{cc}
0     ,   &  0 \leq \alpha < 1/2,\\
-(\ell/2) z^2,  & \alpha = 1/2.
\end{array}
\right.
\end{equation}
\end{Theorem}

See further comments in Section 7.6, Proposition \ref{prop21}.

\begin{Theorem}
\label{thm2}
Suppose that (\ref{6.1}) holds and
$\alpha \in [0,2/3] .$
For each $ R>0 $
there is $n_R >0 $ such that
for $ n\geq n_R  $
the eigenvalues $E_n (z), \; |z| \leq R, $ are
well defined and
\begin{equation}
\label{bb}
E_n (z) = n^2 + z^2 \left (
 \frac{1-2\alpha}{2n^{2-2\alpha}} +
 \frac{\alpha^2-\alpha}{n^{3-2\alpha}}
 +\frac{(1-2\alpha)(8\alpha^2 -14\alpha +3)}{24n^{4-2\alpha}} \right )
 +  O \left( n^{\max (2\alpha-5, 4\alpha-6)} \right ).
\end{equation}
\end{Theorem}
See Theorem~\ref{thm4} in Section 4.4
also.
(For $\alpha = 1/2$
similar formula was given in \cite{BNS}
but it was not correct).

\begin{Theorem}
\label{thm3}
In the case (\ref{6.1}) with $\alpha \in[0, 0.085], $
or $\alpha \in [(2-\sqrt{2})/4, 1/2]$
the Spectral Riemann Surface
$$
G= \{ (\lambda, z)\in \mathbb{C}^2 \;: \;\; \lambda \in Sp(L+zB) \},
$$
is irreducible.
\end{Theorem}
See further comments in Section 7.5, Proposition \ref{prop20}.
In the case of anharmonic oscillator
$$
Ly = -y^{\prime \prime} + x^4 y, \quad By = x^2 y, \quad x \in \mathbb{R}
$$
a question about structure of SRS and its branching points
has been raised and solved (!)
by C. Bender and T. Wu \cite{BW}; see also \cite{S1,Turb1,Turb2,Sh1,Sh2,Sh3}.

The case of  Mathieu--Hill operators
could be deduced to Example (\ref{6.1})+(\ref{6.2});
it has a longer history (see \cite{MS,MSW,B,BC,HG,V1,V2,V4}).
Some observations about Whittaker--Hill operators
could be found in \cite{DM12}, Section 5.4.

4. In the course of proving Theorems \ref{thm1}--\ref{thm3}
we observe a series of facts and inequalities about
the eigenvalues of the operators $L+zB $
which could be of some interest by themselves.
We discuss them in detail in related sections of the paper
or in Section 7.

\section{Localization of the spectra}

1. Well-known methods of Perturbation Theory give information about
the spectra $Sp(L+zB) $ if $L,B \in (\ref{1}), (\ref{2}).$
For a while, let us assume that the sequences $ q, b, c $
satisfy the conditions
\begin{equation}
\label{10.1}
q_k = k^2;
\end{equation}
\begin{equation}
\label{10.2}
|b_k|, |c_k| \leq Mk^\alpha, \quad 0 \leq \alpha < 2.
\end{equation}

For each $n \in \mathbb{N} $ we set
\begin{equation}
\label{10.4}
\Delta_n = \{ z\in \mathbb{C}:\;\;|z| \leq R_n\}, \quad
R_n =n^{1-\alpha}/(8M).
\end{equation}

\begin{Proposition}
\label{prop1} Under the conditions (\ref{10.1}) and (\ref{10.2})
the spectrum of the operator $L+z B$ is discrete, and for each $n$
and $z\in \Delta_n $ there is exactly one eigenvalue $E_n (z) $ in
the strip
$$ H_n = \{\lambda \in \mathbb{C}: \; n^2 -n \leq Re\, \lambda
\leq n^2+n \}.$$

Moreover,
the function $E_n (z) $ is analytic in $\Delta_n, $
\begin{equation}
\label{10.3}
E_n (0) = n^2,
\end{equation}
and
\begin{equation}
\label{10.5}
\left | E_n (z) - n^2 \right | \leq n \quad
\text{if} \quad |z| \leq R_n.
\end{equation}
\end{Proposition}

\begin{proof}
The resolvent--operator
\begin{equation}
\label{11.3}
R_\lambda =(\lambda -L - zB)^{-1} = R^0_\lambda
\left ( 1-zBR^0_\lambda \right )^{-1},
\quad \text{where} \;\; R^0_\lambda = (\lambda - L)^{-1},
\end{equation}
is well defined if
$$ \lambda \not \in Sp (L) \quad \text{and} \quad
|z| \cdot \|BR^0_\lambda\| < 1.
$$

Let
$K_n $ be the open disk with center $n^2 $ and radius $n,$
i.e.,
\begin{equation}
\label{11.6}
K_n =  \{ \lambda \in \mathbb{C}: \;\; |\lambda - n^2| <n \}.
\end{equation}
By (\ref{01}) (see Lemma \ref{lem00} below) we have
$ |z| \cdot \|BR^0_\lambda \| < 1 $ for $|z| \leq R_n $ and
$\lambda \in H_n \setminus K_n ,$
thus
$$ Sp (L+zB) \cap \left ( H_n \setminus K_n \right ) = \emptyset.$$
If $z=0,$ then $Sp (L) = \{k^2:\; k \in \mathbb{N}\},$
so $n^2$ is the only eigenvalue inside the circle $\partial K_n.$
It is simple, and for each $z \in \Delta_n $
the operator $L+zB $ has exactly one simple eigenvalue $E_n (z) \in K_n $
because
$$
\dim \left ( \frac{1}{2\pi i} \int_{\partial K_n}
(\lambda -L-zB)^{-1} d\lambda \right ) \equiv 1.
$$
Moreover, it is well--known that simple eigenvalues
depend analytically on the perturbation parameter
(e.g., see \cite{K}),
and therefore, for each $n,$
$E_n (z) $ is an analytic function on $\Delta_n. $
This completes the proof of Proposition~\ref{prop1}. \vspace{5mm}
\end{proof}

2. The next lemma gives the estimate of the norm $\|BR^0_\lambda
\| .$

\begin{Lemma}
\label{lem00}
Under the assumptions (\ref{10.1}) and (\ref{10.2}),
if $\lambda =x+it \in H_n \setminus K_n $ then
\begin{equation}
\label{01}
\|BR^0_\lambda \| \leq   2M\, \max (2, 2^\alpha ) n^{\alpha-1},
\quad \forall t \in \mathbb{R},
\end{equation}
\begin{equation}
\label{02}
\|BR^0_\lambda \| \leq  2M\, \max (2, 2^\alpha )
n^{\alpha}/|t|, \quad \text{if} \quad
n \leq |t|\leq n^2,
\end{equation}
\begin{equation}
\label{03}
\|BR^0_\lambda \| \leq  4M 2^\alpha |t|^{(\alpha-2)/2},
\quad \text{if} \quad |t|\geq n^2.
\end{equation}
\end{Lemma}

\begin{proof}
Since $R^0_\lambda = \{1/(\lambda-k^2) \} $ is a diagonal operator,
while $B$ is an off--diagonal one, the norm $\|BR^0_\lambda\|$
does not exceed, in view of (\ref{10.2}),
\begin{equation}
\label{11.5}
\|BR^0_\lambda\| \leq \sup_k \frac{|b_k|+|c_{k-1}|}{|\lambda -k^2|}
\leq \sup_k \frac{2Mk^\alpha}{|\lambda -k^2|}.
\end{equation}

For every $t\in \mathbb{R},$
if $k < n,$ then
$|\lambda-k^2| \geq n-1 \geq n/2,$ and therefore,
$k^\alpha/|\lambda -k^2| \leq 2n^{\alpha-1}. $
For $k=n $ we have $ n^\alpha/|\lambda- n^2| \leq n^{\alpha-1} $
because $|\lambda-n^2| \geq n.$
If $n < k \leq 2n, $ then
$ |\lambda -k^2 | \geq k^2 - n^2 -n > n, $
so
$k^\alpha/|\lambda -k^2| \leq 2^\alpha n^{\alpha-1}; $
finally, if $k>2n $ then
$n<k/2,$ and therefore,
\begin{equation}
\label{11.7}
|\lambda - k^2| \geq k^2 - n^2 -n  \geq k^2 -(k/2)^2 - k/2
\geq k^2/2,
\end{equation}
so $k^\alpha/|\lambda -k^2| \leq 2k^{\alpha -2}
\leq  2 n^{\alpha - 2} $
because $\alpha < 2. $
Hence (\ref{01}) holds.

Next we consider the case where $ n\leq |t|\leq n^2. $
Since $|\lambda -k^2| \geq |t| $ we have,
for $k \leq 2n, $ that
$k^\alpha/|\lambda -k^2| \leq (2n)^\alpha/|t|. $
If $k>2n $ then we obtain, as above, that (\ref{11.7}) holds, thus
$$k^\alpha/|\lambda -k^2| \leq 2k^{\alpha}/k^2
\leq 2n^\alpha/n^2 \leq 2n^\alpha/|t|, $$
which proves (\ref{02}).

Consider now the case where $|t| \geq n^2. $
If $ k^2 \leq 4|t|$ then ( since $|\lambda -k^2|\geq |t| $)
$$k^\alpha/|\lambda -k^2| \leq k^\alpha/|t| \leq
2^\alpha |t|^{\alpha/2}/|t|.$$
If $k^2 \geq 4|t| \geq 4n^2 $ then (\ref{11.7}) holds,
thus
$k^\alpha/|\lambda -k^2| \leq 2k^{\alpha -2} \leq 2|4t|^{(\alpha -2)/2},$
which completes the proof of Lemma \ref{lem00}.

\end{proof}

3. By  Proposition \ref{prop1},
for each $k$ there is a disk $\Delta_k $
of radius $R_k (\alpha) = k^{1-\alpha}/(8M) $
with the property that the operator $L+zB$ has exactly one
simple eigenvalue $E_k (z) $
in the strip $ H_k. $
If $ \alpha \in [0,1), $ then $ R_k (\alpha) \uparrow \infty $ as
$k \to \infty.$

Let us fix $ \alpha \in [0,1) $ and   $n \in \mathbb{N}. $
If $ m >n $ then $\Delta_n \subset \Delta_m, $
so for each $ z\in \Delta_n $
$$ Sp (L+zB) \cap
\left (\bigcup_{m\geq n} H_m \right ) \subset \bigcup_{m\geq n}
K_m, $$ where $K_m $ is defined in (\ref{11.6}).
 Set
$$
W_n= \{\lambda \in \mathbb{C}:\; -n < Re \, \lambda < n^2 +n, \;
|Im \, \lambda | < n \}.
$$
\begin{Proposition}
\label{prop2}
Under the conditions (\ref{10.1}), (\ref{10.2}) and (\ref{10.4}), if
$\alpha \in [0,1), $ then for each $z\in \Delta_n $
\begin{equation}
\label{19.3}
Sp(L+zB) \subset W_n \cup \bigcup_{m>n} K_m.
\end{equation}
Moreover, the projector
\begin{equation}
\label{19.5}
P_* (z) = \frac{1}{2\pi i} \int_{\partial W_n}
(\lambda -L-zB)^{-1} d\lambda
\end{equation}
is well defined for $z\in \Delta_n, $ and
\begin{equation}
\label{19.6}
\dim P_* (z) = n.
\end{equation}

\end{Proposition}

\begin{proof} Set
$ H= \{\lambda \in \mathbb{C}: \; Re \, \lambda \leq n^2 +n \}.$
Then
\begin{equation}
\label{19.7}
\sup_k \frac{k^\alpha}{|\lambda - k^2|} = n^{\alpha -1} \quad
\text{for} \quad \lambda \in H\setminus W_n.
\end{equation}
Indeed: if $k\leq n, $ then $|\lambda - k^2| \geq n, $
so $k^\alpha/|\lambda -k^2| \leq n^{\alpha -1};$
if $k>n,$ then $|\lambda - k^2| \geq k, $
thus $k^\alpha/|\lambda -k^2| \leq  k^{\alpha -1} \leq n^{\alpha -1}$
because $\alpha \in [0,1).$

By (\ref{19.7}) and (\ref{11.5}), we obtain that
if $|z| < n^{1-\alpha}/8M,$ then
$$ |z|\cdot \|BR^0_\lambda \| < 1/2 \quad \text{for}
\quad \lambda \in H\setminus W_n.$$
 Therefore, in view of (\ref{11.3}),
for each $z \in \Delta_n, $
$$ Sp(L+zB)\cap (H\setminus W_n) = \emptyset, $$
which proves (\ref{19.3})
because $\mathbb{C}= H\cup \bigcup_{m>n} H_m.$

Moreover, the projector
$$P_* (z) = \frac{1}{2\pi i} \int_{\partial W_n}
(\lambda -L-zB)^{-1} d\lambda.
$$
is well defined for each $z\in \Delta_n, $
and since its dimension is a constant,
we obtain that $\dim P_* (z)= \dim P_* (0)= n. $
\end{proof}

\section{The Taylor coefficients of analytic functions $E_n (z). $}

1. For each $n \in \mathbb{N},$ we consider the rectangles
\begin{equation}
\label{p0}
\Pi = \Pi (n,s) = \{ \lambda \in \mathbb{C}: \; \; |Re \, (\lambda -n^2)|
\leq n, \; |Im\, \lambda | \leq s  \}.
\end{equation}
Then the one-dimensional Riesz projector
\begin{equation}
\label{p1}
P_n (z) = \frac{1}{2 \pi i} \int_{\partial \Pi} (\lambda - L- zB)^{-1}
d \lambda
\end{equation}
is well defined for $|z| \leq R_n $ and does not depend on $s$ for
$s > n+1$ as it follows from (\ref{19.3}) and (\ref{11.6}). The
integrand in (\ref{p1}) is an analytic function of $(\lambda,
z)\in (H_n \setminus \Pi) \times \Delta_n.$

Since
\begin{equation}
\label{p2}
E_n (z)  P_n (z) = \frac{1}{2 \pi i} \int_{\partial \Pi}
\lambda (\lambda - L- zB)^{-1} d \lambda,
\end{equation}
we obtain that
\begin{equation}
\label{p3}
E_n (z) = Trace \left ( \frac{1}{2 \pi i} \int_{\partial \Pi}
\lambda  (\lambda - L- zB)^{-1} d \lambda \right ).
\end{equation}
The formulas (\ref{p1}) -- (\ref{p2}) are basic for what follows
in this section. They are used to derive formulas for
the Taylor coefficients of $E_n (z), $ and to obtain a trace formula.

Let
\begin{equation}
\label{p4}
E_n (z)  = \sum_{k=0}^\infty a_k (n) z^k,  \quad a_0 (n) = n^2,
\end{equation}
be the Taylor expansion of $E_n (z) $ at $0.$

\begin{Proposition}
\label{prop3}
Under the conditions (\ref{10.1}) and (\ref{10.2})  with
$\alpha \in [0,1)$  we have:
\begin{equation}
\label{p5}
a_k (n) = \sum_j
\frac{1}{2 \pi i} \int_{\partial \Pi}
\lambda
\langle R^0_\lambda (BR^0_\lambda)^k e_j, e_j \rangle
d \lambda,
\end{equation}
where
$$ \int_{\partial \Pi}
\lambda
\langle R^0_\lambda (BR^0_\lambda)^k e_j, e_j \rangle
d \lambda = 0  \quad \text{if} \quad |j-n|>k; $$
\begin{equation}
\label{p6}
a_k (n) = \sum_{|j-n|\leq k}
\frac{1}{2 \pi i} \int_{\partial \Pi}
(\lambda -n^2)
\langle R^0_\lambda (BR^0_\lambda)^k e_j, e_j \rangle
d \lambda;
\end{equation}
\begin{equation}
\label{p7}
a_k (n) \equiv 0 \quad \text{for odd} \quad k;
\end{equation}
\begin{equation}
\label{p8}
|a_k (n)| \leq 2(2k+1) \frac{(4M)^k}{n^{(1-\alpha)k-1}}, \quad k\geq 2.
\end{equation}
\end{Proposition}

\begin{proof}
By (\ref{p1}) and (\ref{11.3}),
\begin{equation}
\label{p12}
P_n (z) = \frac{1}{2 \pi i} \int_{\partial \Pi}
\sum_{k=0}^\infty R^0_\lambda (BR^0_\lambda)^k z^k d\lambda =
\sum_{k=0}^\infty p_k (n) z^k,
\end{equation}
where the integrand--series converges absolutely and uniformly for
$ z \in \Delta_n $ and $ \lambda \in \partial \Pi, $
and
\begin{equation}
\label{p13}
p_{k}(n) = \frac{1}{2 \pi i} \int_{\partial \Pi}
R^0_\lambda (BR^0_\lambda)^k  d\lambda, \qquad k=0,1,2, \ldots
\end{equation}
are the Taylor coefficients of $P_n (z) \in (\ref{p1}). $

We have
\begin{equation}
\label{p14}
p_{0} (n) e_n = e_n, \qquad  p_{0}(n) e_j = 0 \quad \text{for} \quad j \neq n.
\end{equation}
Moreover, for each $ k =1,2, \ldots,$
\begin{equation}
\label{p15}
 p_{k}(n) e_j = 0 \quad \text{if} \quad  |j- n| >k.
\end{equation}
Indeed,
\begin{equation}
\label{p16}
p_{k}(n)e_j = \frac{1}{2 \pi i} \int_{\partial \Pi}
R^0_\lambda (BR^0_\lambda)^k e_j d\lambda.
\end{equation}
Since $Be_\nu $ is a linear combination of $e_{\nu -1}$ and
$e_{\nu +1},$  while
$\displaystyle R^0_\lambda e_\nu = \frac{1}{\lambda - \nu^2}\, e_\nu,$
the singularity $\displaystyle \frac{1}{\lambda - n^2} $
(or its power) could appear in the integrand only if $|j-n| \leq k. $
Therefore, if $|j-n| >k, $  then
the integrand is an analytic function on $\Pi, $
so the integral vanishes.

Since $\dim P_n (z) \equiv 1,$
\begin{equation}
\label{p17}
\sum_j \langle P_n (z) e_j, e_j \rangle \equiv 1,
\end{equation}
which implies, in view of (\ref{p14}) and (\ref{p15}), that
\begin{equation}
\label{p18}
\sum_{j} \langle p_{0}(n) e_j, e_j \rangle = 1,
\end{equation}
\begin{equation}
\label{p19}
\sum_{j} \langle p_{k}(n) e_j, e_j \rangle = 0,  \quad
\qquad k=1,2,\ldots.
\end{equation}

Set
\begin{equation}
\label{p21}
E_n (z)  \, P_n (z)  = \sum_{k=0}^\infty d_k (n) z^k.
\end{equation}
Then, by (\ref{p4}) and (\ref{p12}),
\begin{equation}
\label{p22}
  d_k (n) = \sum_{\nu =0}^k a_\nu (n) p_{k-\nu} (n).
\end{equation}
Now (\ref{p18}) and (\ref{p19}) imply,
in view of (\ref{p14}) and (\ref{p15}), that
\begin{equation}
\label{p23}
a_k (n) = \sum_{|j-n|\leq k} \langle d_{k}(n) e_j, e_j \rangle.
\end{equation}
By (\ref{p2}), taking into account the power series expansion of
the resolvent, we obtain
\begin{equation}
\label{p24}
a_k (n) = \sum_{|j-n|\leq k}
\frac{1}{2 \pi i} \int_{\partial \Pi}
\lambda
\langle R^0_\lambda (BR^0_\lambda)^k e_j, e_j \rangle
d \lambda
\end{equation}
Since  $R^0_\lambda $ is a diagonal operator, and
$Be_j $ is a linear combination of $e_{j-1} $ and $e_{j+1},$
we have
\begin{equation}
\label{p24a}
\lambda \langle (BR^0_\lambda)^k e_j, e_j \rangle = 0 \quad
 \forall j \quad \text{ if} \; k \; \text{is odd}.
\end{equation}
Therefore,
the same argument that explains (\ref{p15})
(see (\ref{p16}) and the text after it) shows that the integrals
in (\ref{p24}) are equal to zero if $|j-n|>k, $
which proves (\ref{p5}).

Since $Be_j $ is a linear combination of $ e_{j-1} $ and $e_{j+1}$
and $R^0_\lambda $ is a diagonal operator,
we obtain for odd $k$ that $(BR^0_\lambda)^k e_j $
is a finite linear combination of vectors $e_\nu $ such that $\nu - j$
is odd number, so $\nu \neq j.$
Therefore, if $k$ is odd, then for each $j$
the integrands in (\ref{p5}) are equal to zero, which proves (\ref{p7}).

By (\ref{p15}), (\ref{p16}) and (\ref{p19}) we have
$$ \sum_{|j-n| \leq k}
\frac{1}{2 \pi i} \int_{\partial \Pi}
\langle R^0_\lambda (BR^0_\lambda)^k e_j, e_j \rangle d\lambda = 0,
$$
thus (\ref{p5}) implies (\ref{p6}).

Next we prove (\ref{p8}). Let us replace the contour $\partial \Pi
$ in (\ref{p6}) by the circle $\partial K_n =\{\lambda: \;
|\lambda-n^2|=n \}. $ Fix $j$ with $|j-n| \leq k $ and consider
the corresponding integral. The integrand does not exceed
$$ \sup_{\lambda \in \partial K_n}
\left ( |\lambda -n^2| \cdot \|R^0_\lambda \| \cdot
\|BR^0_\lambda \|^k \right ).
$$
By (\ref{01}),
we have, for $\alpha \in [0,1),$
$$ \|BR^0_\lambda \| \leq 4Mn^{\alpha -1} \quad \text{if} \quad
\lambda \in \partial K_n.
$$
On the other hand $|\lambda -n^2 | =n $ on $\partial K_n, $ and

$$
\|R^0_\lambda \| =\sup_j \frac{1}{\lambda - j^2} \leq \frac{2}{n}
\quad \text{for} \quad \lambda \in \partial K_n.
$$
Thus, for each $j,$
the integrand does not exceed $2(4M)^k n^{(\alpha -1)k} $
and the length of $\partial K_n $ is equal to $2 \pi n, $
which leads to the estimate (\ref{p8}).
\end{proof}

2. Next we give another integral representation of the coefficients
$a_k (n). $

\begin{Proposition}
\label{prop4}
Under the conditions (\ref{10.1}) and (\ref{10.2}) with
$\alpha \in [0,1) $  we have, for each $ k \geq 2, $
\begin{equation}
\label{p27}
a_k (1) = \varphi_k (1), \quad
a_k (n) = \varphi_k (n) - \varphi_k (n-1),   \quad n\geq 2,
\end{equation}
with
\begin{equation}
\label{p28}
\varphi_k (n) = \sum_{j}
\frac{1}{2 \pi i} \int_{h_n}
\lambda
\langle R^0_\lambda (BR^0_\lambda)^k e_j, e_j \rangle
d \lambda,
\end{equation}
where
\begin{equation}
\label{p29}
h_n = \{\lambda \in \mathbb{C}: \;\; Re \, \lambda = n^2 +n\},
\end{equation}
and
$$ \int_{h_n} \lambda
\langle R^0_\lambda (BR^0_\lambda)^k e_j, e_j \rangle
d \lambda =0  \quad \text{if} \quad |j-n|>k. $$
Moreover,
\begin{equation}
\label{p29a}
|\varphi_k (n)| \leq \frac{C_k}{n^{(1-\alpha)k-2}},
\quad k> 1, \;\; C_k = (2k+1)(8M)^k.
\end{equation}
\end{Proposition}

\begin{proof}
Letting $s \to \infty $ in (\ref{p5})
we obtain (\ref{p27})--(\ref{p29}).
To justify this limit procedure, we have to explain that

(i) the integrals over $h_n $ and $h_{n-1} $ converge;

(ii) the integrals over horizontal sides
of $\partial \Pi (n,s) \in (\ref{p0})$
go to zero as $n \to \infty;$

(iii) the integrals over $h_n $ are equal to zero if
$|j - n|>k; $

(iv) $a_k (1) = \varphi_k (1).$

Indeed, (i) and (ii) hold because  the integrand in (\ref{p5}),
for each even $k \geq 2, $ is a linear combination of rational
functions of the form
\begin{equation}
\label{p30}
Q(J, \lambda)=
\frac{\lambda }{(\lambda -j_0^2)(\lambda -j_1^2)\cdots (\lambda - j_k^2)},
\quad J= (j_0, \ldots, j_k),
\end{equation}
and therefore, the integrand decays faster than $1/|\lambda|^2 $
as $ |\lambda| \to \infty. $

(iii) If $ j - n >k $  (respectively  $n-j >k $),
then the integrand is a sum of terms (\ref{p30})
with $ j_0, \ldots, j_k > n $
(respectively $ j_0, \ldots, j_k < n $ ).
Consider the contour that
consist of the segment  $\{\lambda \in h_n : \; |Im \, \lambda| \leq s\}  $
and the left half (respectively right half)
of the circle with center $n^2 +n $ and
radius $s.$ Since the integrand is an analytic function inside the
contour, the integral is equal to zero. Letting $s \to \infty $
we obtain that the integral over $h_n $ is zero, because
the integral over the half--circle goes to zero due to the fact that
the integrand decays as $1/|\lambda|^2$ or more rapidly.

The same argument shows, for each $j,$ that the integral over the
imaginary line $ Re \lambda = 0 $ equals zero, which explains
(iv).

Finally, we prove (\ref{p29a}).
By (\ref{p28}), the function $\varphi_k (n) $ is a sum of
at most $2k+1 $ integrals over $h_n $ of the form
\begin{equation}
\label{011}
\frac{1}{2 \pi i} \int_{h_n}
\lambda
\langle R^0_\lambda (BR^0_\lambda)^k e_j, e_j \rangle
d \lambda.
\end{equation}
The absolute value of the integral (\ref{011}) does not exceed
\begin{equation}
\label{012}
\frac{1}{2 \pi } \int_\mathbb{R} F(t) dt,  \quad
\text{where} \quad
F(t) = \|\lambda  R^0_\lambda
 (BR^0_\lambda)^k \|, \;\; \lambda = n^2 +n +it.
\end{equation}
Next we estimate from above
$F(t)  \leq \|\lambda R^0_\lambda \| \cdot \|B  R^0_\lambda \|^k.$
Lemma~\ref{lem00} gives estimates of the norm
$\|B  R^0_\lambda \| $ on each of the three sets
$$ I_1 = \{t: \; |t|\leq n\}, \quad
I_1 = \{t: \; n \leq |t|\leq n^2\}, \quad
I_1 = \{t: \; |t|\geq n^2\}. $$

On the other hand we have
\begin{equation}
\label{013}
\|\lambda R^0_\lambda \| = \frac{|n^2+n+it|}{|n+it|} \leq
\left \{
\begin{array}{cc}
n+1, &  t\in I_1\\
2n^2/|t|, & t\in I_2\\
2,  &  t \in I_3
\end{array}
\right.
\end{equation}
If we combine (\ref{013}) with the estimates (\ref{01})--(\ref{03})
from Lemma \ref{lem00} we get
\begin{equation}
\label{014}
F(t) \leq
\left \{
\begin{array}{cc}
2n \left ( 4Mn^{\alpha -1} \right )^k, & t\in I_1\\
2n^2 \left ( 4Mn^{\alpha} |t|^{-1} \right )^k,  & t\in I_2\\
2 \left ( 8M  |t|^{(\alpha -2)/2} \right )^k,  &  t \in I_3
\end{array}
\right.
\end{equation}
Therefore, since
$$
\int_{\mathbb{R}} F(t) =\int_{I_1} F(t) + \int_{I_2} F(t)+
\int_{I_3} F(t),
$$
the estimates (\ref{014}) imply that (\ref{p29a}) holds.

\end{proof}

3. The formulas (\ref{p27}) and (\ref{p28}) could be used to find
the Taylor coefficients of $E_n (z).$
Indeed,
under the conditions (\ref{10.1}) and (\ref{10.2}) with
$\alpha \in [0,1), $
a computation based on the standard residue approach shows that
\begin{equation}
\label{p31a}
\varphi_2 (n) = - \frac{b_n c_n}{2n+1},
\end{equation}
\begin{equation}
\label{p32a}
\varphi_4 (n) = \frac{b_n^2 c_n^2}{(2n+1)^3} -
\frac{b_n b_{n+1}c_n c_{n+1}}{(2n+1)^2(4n+4)}-
\frac{b_n b_{n-1}c_n c_{n-1}}{4n(2n+1)^2}.
\end{equation}

For any off--diagonal sequences $b,c \in (\ref{10.1}) +
(\ref{10.2}),$ it follows from (\ref{p29a}) that as $n\to \infty $
\begin{equation}
\label{p32b}
\varphi_k (n) \to 0  \quad \text{if} \;\;
\alpha < 2/3, \; k\geq 6,
\end{equation}
and by (\ref{p31a}), (\ref{p32a})
\begin{equation}
\label{p33b}
\varphi_2 (n) \to 0,   \quad \text{if} \;\; \alpha < 1/2, \quad
\varphi_4 (n) \to 0,   \quad \text{if} \;\; \alpha < 3/4.
\end{equation}
Now, by (\ref{p32b}) and (\ref{p27}),
\begin{equation}
\label{p36a}
\sum_{n=1}^\infty a_k (n) = 0 \quad \text{if} \quad
k\geq 6, \;\; \alpha \in [0,1/2].
\end{equation}

If (\ref{6.1}) holds,
then
\begin{equation}
\label{p31}
\varphi_2 (n) = - \frac{n^{2\alpha}}{2n+1},
\end{equation}
\begin{equation}
\label{p32}
\varphi_4 (n) = \frac{n^{4\alpha}}{(2n+1)^3} -
\frac{n^{2\alpha}(n+1)^{2\alpha}}{(2n+1)^2(4n+4)}-
\frac{(n-1)^{2\alpha}n^{2\alpha}}{4n(2n+1)^2}.
\end{equation}

By (\ref{p27}) and (\ref{p31}) we obtain
\begin{equation}
\label{p33}
a_2 (1) =\varphi_2 (1) = -\frac{1}{3}, \quad
a_2 (n)  =  \frac{(n-1)^{2\alpha}}{2n-1}- \frac{n^{2\alpha}}{2n+1}
\quad \text{for} \quad n\geq 2.
\end{equation}
Observe that $\varphi_2 (n) \to 0 $ if $ \alpha \in [0, 1/2), $
while  $\varphi_2 (n) \to -1/2 $ if $\alpha = 1/2. $
Thus we have
\begin{equation}
\label{p34}
\sum_{n=1}^\infty a_2 (n) = 0 \quad \text{for} \quad \alpha \in [0, 1/2)
\end{equation}
and
\begin{equation}
\label{p35}
\sum_{n=1}^\infty a_2 (n) = - \frac{1}{2} \quad \text{if} \quad \alpha = 1/2.
\end{equation}

By (\ref{p32}) we obtain that
$\varphi_4 (n) \to 0 $ if $ \alpha \in [0,1/2],$
so (\ref{p27}) yields
\begin{equation}
\label{p36}
\sum_{n=1}^\infty a_4 (n) = 0 \quad \text{if} \quad \alpha \in [0,1/2].
\end{equation}

\section{Asymptotics of $E_n (z)$}

In this section we study the asymptotic behavior of
$E_n (z) $ for large $n.$
Our approach is based on the fact that
the eigenvalue function $E_n (z) $ satisfies a quasi-linear equation.
Of course, the same estimates and formulas
could be found if one follows the Raleigh--Schr\"odinger scheme with
recurrences for the Taylor coefficients
$$
\lambda (z) = \sum_{k=0}^\infty a_{2k} (n) z^{2k}, \quad a_0 (n) = n^2,
$$
$$
f(z) = \sum_{j=0}^\infty f_j z^j, \quad f_j \in \ell^2 (\mathbb{N}), \;
f_0 = e_n,
$$
as they would come if one substitute the above formulas into
(\ref{b1}).

\vspace{3mm}

1. Throughout this section we assume that (\ref{10.1}) and
(\ref{10.2}) with $\alpha \in [0,1/2] $  hold, but after
(\ref{b10}) we assume that (\ref{6.1}) holds also.

Suppose that $n $ and $z\in \Delta_n $ are fixed
and $\lambda = E_n (z) $ is the corresponding eigenvalue of the operator
$L+zB.$ Then we have
\begin{equation}
\label{b1}
(L+zB)f = \lambda f
\end{equation}
for some $f\neq 0.$
Let $P $ be the projector defined by $Px = \langle x,e_n \rangle e_n,$
and let $Q= 1-P.$
The equation (\ref{b1})
is equivalent to the system of two equations
\begin{equation}
\label{b2}
(\lambda -L) f_1 = zPB (f_1 + f_2),
\end{equation}
\begin{equation}
\label{b3}
(\lambda -L) f_2 = zQB (f_1 + f_2),
\end{equation}
where $ f_1 = Pf, \; f_2 = Qf. $
The operator $\lambda - L $ is invertible on the range of the projector
$Q; $ we set
\begin{equation}
\label{b5}
De_k = \frac{1}{\lambda - k^2} e_k \quad \text{if} \quad k\neq n,
\qquad D e_n = 0.
\end{equation}
Then $D $ is well defined in $\ell^2, $
and $ (\lambda - L) Dx = x  $
on the range of $Q.$

Acting on both sides of (\ref{b3}) by the operator $BD $
we obtain
\begin{equation}
\label{b6}
Bf_2 = zTBf_1 + zT Bf_2,
\end{equation}
where
\begin{equation}
\label{b7}
T= BD.
\end{equation}
The operator $1-zT $ is invertible for each $z \in \Delta_n. $
Indeed, since $Te_k = BR^0_\lambda e_k $ for $k \neq n $
and $Te_n = 0, $ the proof of
(\ref{01}) shows that
\begin{equation}
\label{b7a}
\|T \|\leq 4M \cdot
n^{\alpha-1} \quad \text{for} \quad
\lambda \in H_n.
\end{equation}
Thus we have
$$ \|zT\| \leq |z|\cdot \|T\|
< 1 $$
for each $z \in \Delta_n $ and each $\lambda \in H_n. $

Solving (\ref{b6}) for $Bf_2 $ we obtain
\begin{equation}
\label{b8}
Bf_2 = z(1-zT)^{-1} TBf_1.
\end{equation}
Inserted into (\ref{b2}), this leads to
$$
(\lambda -L)f_1 = zPBf_1 + z^2 P(1-zT)^{-1} TBf_1,
$$
which implies
(since $1+ zT(1-zT)^{-1} = (1-zT)^{-1} $)
\begin{equation}
\label{b9}
(\lambda -L)f_1 = z P(1-zT)^{-1} Bf_1,
\end{equation}
where $f_1= const \cdot e_n \neq 0 $ (otherwise, by (\ref{b3}) it follows
that $f_2 = 0, $ so $f= f_1+f_2 =0,$ which contradicts $f\neq 0$).
Since $Le_n = n^2 e_n,$
the equation (\ref{b9}) is equivalent to
\begin{equation}
\label{b10}
\lambda -n^2 = z \langle (1-zT)^{-1} Be_n,e_n \rangle.
\end{equation}

Since
\begin{equation}
\label{b11}
Be_k = (k-1)^\alpha e_{k-1} + k^\alpha e_{k+1},
\end{equation}
we have, by (\ref{b5}) and (\ref{b7}), that
\begin{equation}
\label{b12}
Te_k = \frac{1}{\lambda - k^2}
\left ( (k-1)^\alpha e_{k-1} + k^\alpha e_{k+1} \right ), \quad Te_n = 0,
\end{equation}
and therefore,
\begin{equation}
\label{b13}
\langle T^{2k} B e_n, e_n \rangle =0, \quad k=0, 1, 2, \ldots.
\end{equation}

Let
$$E_n (z) = n^2 + a_1 (n) z + a_2 (n) z^2 + \cdots $$
be the Taylor expansion of $E_n (z). $
Set for convenience
\begin{equation}
\label{b14}
\zeta (z) =  E_n (z) - n^2 = a_1 (n) z + a_2 (n) z^2 + \cdots.
\end{equation}
Then, by (\ref{b10}),
\begin{equation}
\label{b15}
\zeta (z) = \langle TBe_n, e_n \rangle z^2 +
\langle T^3 Be_n, e_n \rangle z^4 +\langle T^5 Be_n, e_n \rangle z^6
+ \cdots,
\end{equation}
where, by (\ref{b12}), the operator $T$ depends rationally
on $\lambda = E_n (z) = \zeta (z) + n^2. $
It is easy to see, by induction, that (\ref{b15}) yields
$a_{2k+1}(n) =  0, \; k\in \mathbb{N}$
(in fact, we know this from Section 3.1, see (\ref{p7})).
Thus we have
\begin{equation}
\label{b16}
\zeta (z) = a_2 (n) z^2 + a_4 (n) z^4 + \cdots.
\end{equation}

One may use (\ref{b15}) to compute the Taylor coefficients of $\zeta (z).$
Indeed, the right side of (\ref{b15})
is a power series in $z$ which coefficients
are rational functions of
$\lambda = \zeta + n^2 $ without a singularity at 0.
So, replacing these rational functions with their power series expansions
at 0, and replacing $\zeta $ with its power expansion (\ref{b14}),
we obtain (comparing the resulting power series expansion
on the left and on the right)
a system of equations for the coefficients
$a_2 (n), a_4 (n), \ldots. $

Next we compute some of these coefficients. By
(\ref{b11})--(\ref{b15}) it follows that
\begin{equation}
\label{0b}
\zeta = z^2
\left ( \frac{(n-1)^{2\alpha}}{2n-1 +\zeta}
-\frac{n^{2\alpha}}{2n+1 - \zeta} \right )  +
\end{equation}
$$
z^4
\left ( \frac{(n-1)^{2\alpha}(n-2)^{2\alpha}}{(2n-1 +\zeta)^2 (4n-4+\zeta)}
-\frac{n^{2\alpha}(n+1)^{2\alpha}}{(2n+1 - \zeta)^2(4n+4+\zeta)} \right )
 + \cdots
 $$
$$= z^2
\left ( \frac{(n-1)^{2\alpha}}{2n-1} \left
(1-\frac{\zeta}{2n-1}+\cdots \right) -\frac{n^{2\alpha}}{2n+1}
\left (1+\frac{\zeta}{2n+1}+ \cdots \right) \right )+
$$
$$  z^4
\left [ \frac{(n-1)^{2\alpha}(n-2)^{2\alpha}}
{(2n-1)^2 (4n-4)}\left (1-\frac{\zeta}{2n-1}+\cdots \right)^2
\left (1-\frac{\zeta}{4n-4}+\cdots \right) \right. $$
$$
\left. -\frac{n^{2\alpha}(n+1)^{2\alpha}}{(2n+1)^2(4n+4)}
\left (1+\frac{\zeta}{2n+1}+\cdots \right)^2
\left (1+\frac{\zeta}{4n+4}+\cdots \right)
\right ] + \cdots
$$
Hence we obtain
\begin{equation}
\label{b20}
a_2 (\alpha,n) = \frac{(n-1)^{2\alpha}}{2n-1} - \frac{n^{2\alpha}}{2n+1},
\quad n\geq 2;
\end{equation}
\begin{equation}
\label{b21}
a_4 (\alpha,n) = (-a_2 (n))
\left ( \frac{(n-1)^{2\alpha}}{(2n-1)^2} +
 \frac{n^{2\alpha}}{(2n+1)^2} \right )
+
\end{equation}
$$ +
\frac{(n-1)^{2\alpha}(n-2)^{2\alpha}}{(2n-1)^2(4n-4)} -
\frac{n^{2\alpha}(n+1)^{2\alpha}}{(2n+1)^2(4n+4)}, \quad n\geq 3.
$$
The same method gives
\begin{equation}
\label{b22}
a_6 (\alpha,n) = \sigma_1 (n) -a_2(n) \sigma_2 (n)  -a_4 (\alpha,n)
\sigma_3 (n), \quad n\geq 4,
\end{equation}
where
\begin{equation}
\label{b23}
\sigma_1 (n) = \frac{(n-1)^{2\alpha}(n-2)^{4\alpha}}{(2n-1)^3(4n-4)^2}
-\frac{n^{2\alpha}(n+1)^{4\alpha}}{(2n+1)^3 (4n+4)^2}
\end{equation}
$$
+ \frac{(n-1)^{2\alpha}(n-2)^{2\alpha}(n-3)^{2\alpha}}
{(2n-1)^2(4n-4)^2(6n-9)}
- \frac{n^{2\alpha}(n+1)^{2\alpha}(n+2)^{2\alpha}}
{(2n+1)^2(4n+4)^2(6n+9)};
$$
\begin{equation}
\label{b24}
\sigma_2 (n) = \frac{(n-1)^{2\alpha}(n-2)^{2\alpha}}{(2n-1)^2(4n-4)}
\left ( \frac{2}{2n-1} + \frac{1}{4n-4} \right ) +
\end{equation}
$$
+ \frac{n^{2\alpha}(n+1)^{2\alpha}}{(2n+1)^2(4n+4)}
\left ( \frac{2}{2n+1} + \frac{1}{4n+4} \right ) +
\frac{n^{2\alpha}}{(2n+1)^3} -\frac{(n-1)^{2\alpha}}{(2n-1)^3} ;
$$
\begin{equation}
\label{b25}
\sigma_3 (n) =
\frac{(n-1)^{2\alpha}}{(2n-1)^2} + \frac{n^{2\alpha}}{(2n+1)^2}.
\end{equation}

Of course, the case of small $n$ requires a special treatment.
For example, if $n=1,$ then
with
$$
\zeta = a_2(1) z^2 + a_4 (1) z^4 + a_6 (1) z^6 +\cdots
$$
we have
$$
\zeta = z^2 \left ( \frac{1}{\zeta -3} \right ) +
z^4 \left ( \frac{2^{2\alpha}}{(\zeta -3)^2(\zeta-8)} \right )+
$$
$$
z^6 \left ( \frac{2^{4\alpha}}{(\zeta -3)^3(\zeta-8)^2}+
\frac{2^{2\alpha}3^{2\alpha}}{(\zeta -3)^2(\zeta-8)^2(\zeta -15)} \right )
+ \cdots,
$$
which leads to
\begin{equation}
\label{25b}
a_2 (1) = -\frac{1}{3}, \qquad a_4 (1) = \frac{1}{27} -
\frac{2^{2\alpha}}{72}
\end{equation}
(compare with (\ref{p27}), (\ref{p31}), (\ref{p32})), and
\begin{equation}
\label{26b}
a_6 (1) = -\frac{2^{4\alpha}}{3^3\cdot 8^2}-
\frac{2^{2\alpha}3^{2\alpha}}{3^3\cdot 8^2\cdot 5} +
\frac{2^{2\alpha}}{3\cdot 8^2} -\frac{2}{3^5}.
\end{equation}

2. The following lemma gives the asymptotic behavior of $a_{2k} (\alpha,n)$
as $n \to \infty.$

\begin{Lemma}
\label{lem31}
Under the condition (\ref{6.1}), if
$\alpha \in [0,1), $ then
\begin{equation}
\label{b26}
a_{2k} (\alpha,n) = O( n^{2k(\alpha -1)}).
\end{equation}
\end{Lemma}

\begin{proof}
We prove (\ref{b26}) by induction in $k.$
If $k=1,$ then (\ref{b20}) yields
\begin{equation}
\label{b27}
a_2 (\alpha,n) = O( n^{2(\alpha -1)}).
\end{equation}
If $k=2,$ then (\ref{b21}) gives $a_4 (\alpha,n) $ as a sum of two
expressions. For the first one we obtain, in view of (\ref{b27}), that
$$
a_2  (\alpha,n) \left ( \frac{(n-1)^{2\alpha}}{(2n-1)^2} +
 \frac{n^{2\alpha}}{(2n+1)^2} \right ) =
 O( n^{2(\alpha -1)})\cdot O( n^{2(\alpha -1)}) =O( n^{4(\alpha -1)}).
 $$
The remaining part of (\ref{b21}) is
\begin{equation}
\label{b28}
\frac{(n-1)^{2\alpha}(n-2)^{2\alpha}}{(2n-1)^2(4n-4)} -
\frac{n^{2\alpha}(n+1)^{2\alpha}}{(2n+1)^2(4n+4)}.
\end{equation}
Each term of this difference is $O(n^{4\alpha -3}).$
But (\ref{b28}) is $O( n^{4(\alpha -1)})$
due to the Mean Value Theorem. Indeed,
let $f(m) = m^{2\alpha}(m+1)^{2\alpha}(4m+4)^{-1} $ and
$g(m) = (2m+1)^{-2}.$  Then
(\ref{b28}) may be written as
$$
f(n-2)g(n-1)-f(n)g(n) = (f(n-2)- f(n))g(n-1) + f(n) (g(n-1)-g(n)).
$$
Since
$$f^\prime (t)  = O(f(n)/n), \quad
g^\prime (t)  = O(g(n)/n), \quad  \text{for} \quad t\in[n-2,n],
$$
by the Mean Value Theorem the expression (\ref{b28}) is
$O(n^{4\alpha -4}) $ which proves (\ref{b26}) for $k=2.$

Fix $k \geq 3 $ and assume that (\ref{b26}) holds for
$1, \ldots, k-1.$
Then by (\ref{b15}) and (\ref{b16}) we obtain, in view of
(\ref{0b}),
that
$$
a_{2k} =\langle T^{2k-1}Be_n, e_n \rangle +
\sum C_{m_1\ldots m_{k-1}} a_2^{m_1} \cdots a_{2(k-1)}^{m_{k-1}},
$$
where $m_1 + 2m_2 +\cdots + (k-1) m_{k-1} = k, $ and
$$
T= BD \quad \text{with} \quad De_n = 0, \;\; De_\nu =
\frac{1}{n^2-\nu^2} e_\nu.
$$
In addition,
for each term of the sum, we have
$$
 C_{m_1\ldots m_{k-1}} a_2^{m_1} \cdots a_{2(k-1)}^{m_{k-1}} =
 O(n^{2k(\alpha -1)}).
$$
(See (\ref{b22})--(\ref{b25}) for the case $k=3.$)
Thus Lemma \ref{lem31} will be proved if we show that
\begin{equation}
\label{b29}
 \langle T^{2k-1}Be_n, e_n \rangle =
O ( n^{2k(\alpha -1)}).
\end{equation}

Set
\begin{equation}
\label{2b}
B = B_{+1} + B_{-1} \quad \text{and} \quad T =T_{+1} + T_{-1},
\end{equation}
where
\begin{equation}
\label{3b}
B_{+1} e_k = k^\alpha e_{k+1}, \qquad  B_{-1}e_k = (k-1)^\alpha e_{k-1},
\end{equation}
and
\begin{equation}
\label{4b}
T_{+1} = B_{+1}D, \qquad   T_{-1} =B_{-1}D.
\end{equation}
Then
\begin{equation}
\label{5b}
\langle T^{2k-1}Be_n, e_n \rangle = \sum_{\varepsilon} \omega (\varepsilon),
\end{equation}
where the summation is over all $2k$--tuples
$\varepsilon = (\varepsilon_1, \ldots, \varepsilon_{2k})$
with $\varepsilon_\nu = \pm 1, $ and
\begin{equation}
\label{6b}
\omega (\varepsilon) = \langle T_{\varepsilon_{2k-1}}\cdots
T_{\varepsilon_2}B_{\varepsilon_1} e_n, e_n \rangle.
\end{equation}

Let
\begin{equation}
\label{7b}
\delta (\varepsilon) = (\delta_1, \ldots, \delta_{2k}), \quad
\delta_\nu = \delta_\nu (\varepsilon) = \varepsilon_1 +
\cdots +\varepsilon_\nu, \;\; \nu = 1, \ldots, 2k;
\end{equation}
then $T_{\varepsilon_\nu} \cdots T_{\varepsilon_2}B_{\varepsilon_1} e_n =
const \cdot e_{n+\delta_\nu}. $
Therefore,
since $De_n =0,$ we have
$\omega (\varepsilon) \neq 0 $ if and only if $\delta_{2k} =0$
and
$\delta_\nu \neq 0 $ for $\nu \neq 2k.$

Now (\ref{5b}) implies that
\begin{equation}
\label{8b}
 \langle T^{2k-1}Be_n, e_n \rangle =
 \sum_{\varepsilon \in e^+}
 [ \omega (\varepsilon) + \omega (-\varepsilon) ],
\end{equation}
where the summation is over the set $e^+ $ of all $2k$-tuples
$\varepsilon $ such that
$\delta_nu (\varepsilon )  > 0 $ for $ \nu =1, \ldots, 2k-1.$
Since the cardinality of $e^+ $ does not exceed $2^{2k}, $
(\ref{b29}) will be proved if we show,
for each $\varepsilon \in e^+, $ that
\begin{equation}
\label{9b}
\omega (\varepsilon)  +\omega (-\varepsilon)= O (n^{2k(\alpha -1)}).
\end{equation}
By (\ref{2b})--(\ref{4b}) we obtain
$$
\omega (\varepsilon) = -
\frac{\prod_{\nu=1}^{2k} (n+ \delta_{\nu -1}+
(\varepsilon_\nu -1 )/2 )^\alpha }
{\prod_{\nu=1}^{2k-1} \left (\delta_\nu (2n + \delta_\nu) \right )},
\quad \delta_0 = 0, \; \delta_\nu = \varepsilon_1 +\cdots + \varepsilon_\nu.
$$
Now, as above, the Mean Value Theorem
may be used to show that
(\ref{9b}) holds. This completes the proof of Lemma \ref{lem31}.

\end{proof}

3. {\em Proof of Theorem \ref{thm2}.}
By Proposition \ref{prop1} we know that, with $R_n=n^{1-\alpha}/(8M),$
$$| E_n (z)- n^2| \leq n  \quad \text{for}
\quad z \in \Delta_n = \{\zeta : \; |\zeta| \leq R_n \}. $$

\begin{Lemma}

\label{lem12}
For each $k=1,2,\ldots ,$
\begin{equation}
\label{45.1}
|a_k (n) |  =  \frac{1}{k!} \left |E_n^{(k)} (0) \right |
\leq  (8M)^k n^{1-k(1-\alpha)}.
\end{equation}
\end{Lemma}

\begin{proof}
Indeed,
$E_n (z) $ is analytic in $\Delta_n. $
Therefore, the Cauchy inequality for the Taylor coefficients
of $E_n (z) $ at 0 gives (\ref{45.1}).
\end{proof}

Now, for $|z| \leq R,$ we obtain
\begin{equation}
\label{b34}
|E_n (z) - n^2 -\sum_{k=1}^6 a_{2k}z^{2k} |
\leq \sum_{k=7}^\infty
|a_{2k} (n)| R^{2k} \leq \frac{C_n}{n^{13-14\alpha}},
\end{equation}
where $C_n= (8MR)^{14} \sum_{k\geq 0} (8MR)^{2k}/n^{2k(1-\alpha)}$
is a bounded sequence.

On the other hand (\ref{b20}) and (\ref{b21}) imply that
\begin{equation}
\label{b35}
a_2 (\alpha,n) =
 \frac{(1-2\alpha)}{2n^{2-2\alpha}} +
 \frac{(\alpha^2-\alpha)}{n^{3-2\alpha}} +
 \frac{(1-2\alpha)(8\alpha^2 -14\alpha +3)}{24 n^{4-2\alpha}}
+ O(n^{2\alpha-5})
\end{equation}
and
\begin{equation}
\label{b36}
a_4 (\alpha,n) =  O (n^{4\alpha-6}).
\end{equation}
The formulas (\ref{b22})--(\ref{b25}) yield
\begin{equation}
\label{b37}
a_6 (\alpha,n) =  O (n^{6\alpha-10}).
\end{equation}
Analogous computations show that
\begin{equation}
\label{b38}
a_8 (\alpha,n) =  O (n^{8\alpha-14}).
\end{equation}
Finally, by Lemma \ref{lem31}
\begin{equation}
\label{b39}
a_{10} (\alpha,n) =  O (n^{10\alpha-10}), \quad
a_{12} (\alpha,n) =  O (n^{12\alpha-12}).
\end{equation}

Now (\ref{b34})--(\ref{b39}) imply (\ref{bb}).
Indeed, if $\alpha \in [0,1/2],$
then $2\alpha - 5 \geq 4\alpha -6;$
moreover,
$$12\alpha - 12 \leq 10\alpha - 10 \leq 2\alpha - 5 $$
and
$14\alpha - 13 \leq 2\alpha - 5, $ thus
(\ref{bb}) holds.

If $\alpha \in [1/2,2/3],$ then
$2\alpha - 5 \leq 4\alpha -6;$
so, since
$$12\alpha - 12 \leq 10\alpha - 10 \leq 4\alpha - 6 $$
and
$14\alpha - 13 \leq 4\alpha - 6, $
we obtain that (\ref{bb}) holds.
This completes the proof of Theorem \ref{thm2}.

4. We consider separately the case where $\alpha = 1/2 $
in the following theorem.

\begin{Theorem}
\label{thm4}
If $|z| \leq R, $  then
\begin{equation}
\label{b50}
E_n (1/2, z) = n^2 - \frac{z^2}{4n^2} -
\frac{2z^2+3z^4}{32n^4} + O(1/n^6).
\end{equation}
\end{Theorem}

\begin{proof}
If $\alpha = 1/2,$ then (\ref{b34}) implies
\begin{equation}
\label{b54}
|E_n (z) - n^2 -\sum_{k=1}^6 a_{2k}z^{2k} |
\leq \sum_{k=7}^\infty
|a_{2k} (n)| R^{2k} \leq \frac{C_n}{n^6},
\end{equation}
where $C_n= (8MR)^{14} \sum_{k\geq 0} (8MR)^{2k}/n^k$
is a bounded sequence.
On the other hand, from (\ref{b20})--(\ref{b25}) it follows that
\begin{equation}
\label{b56}
a_2  = -\frac{1}{4n^2 -1},
\end{equation}
\begin{equation}
\label{b57}
a_4  = \frac{1}{4(2n+1)^3} -\frac{1}{4(2n-1)^3}
\end{equation}
\begin{equation}
\label{b58}
a_6  =
- \frac{1}{(2n+3)(2n+1)^5(2n-1)} +\frac{1}{(2n+1)(2n-1)^5(2n-3)}.
\end{equation}
The same approach that leads to (\ref{b20})--(\ref{b25}) gives
\begin{equation}
\label{b59}
a_8  = \frac{-327-16080n^2-63136n^4+29440n^6 +39168n^8}
{32(n-1)(n+1)(2n-3)(2n+3)(2n-1)^7(2n+1)^7}.
\end{equation}
and
\begin{equation}
\label{b60}
a_{10}  = \frac{3915 +280676n^2 +2496992 n^4 +
2635904 n^6 - 3111168 n^8 -1158144 n^{10}}
{8(n-1)(n+1)(2n-3)(2n+3)(2n-5)(2n+5)(2n-1)^9 (2n+1)^9}.
\end{equation}

By (\ref{b57})--(\ref{b60}) we obtain
\begin{equation}
\label{b61}
a_2(1/2,n)= -\frac{1}{4n^2} -\frac{1}{16n^4} +O(n^{-6}),
\end{equation}
\begin{equation}
\label{b62}
a_4(1/2,n)=  -\frac{3}{32n^4} +O(n^{-6}),
\end{equation}
and
\begin{equation}
\label{b64}
a_6(1/2,n)=O(n^{-8}),\quad a_8(1/2,n)=O(n^{-10}),\quad
a_{10}(1/2,n)=O(n^{-14}).
\end{equation}
In addition, Lemma \ref{lem31} implies that
\begin{equation}
\label{b65}
a_{12}(1/2,n)=O(n^{-6}).
\end{equation}
Now
(\ref{b50}) follows from (\ref{b54}) and
(\ref{b61})--(\ref{b65}).

\end{proof}

\begin{Remark}
\label{rem1}
\end{Remark}
We evaluate $a_{12} $ in (\ref{b65}) by using
the general estimate (\ref{b26}) from
Lemma \ref{lem31}. However, {\it a direct computation
of the coefficients $a_{2k} (n)$ for $ 6\leq k \leq 14$
shows that each of them is} $O(1/n^{16}).$
Estimating the remainder as in the proof of Theorem \ref{thm4},
we get
\begin{equation}
\sum_{k\geq 15} a_{2k}(n) z^k  =  O(1/n^{14}), \quad  |z| \leq R.
\end{equation}
So, by (\ref{b60}), we have
\begin{equation}
E_n (z)= n^2 + a_2 z^2 + a_4 (n) z^4 + a_6 (n) z^6 + a_8 (n) z^8
+ O(1/n^{14}), \quad |z| \leq R.
\end{equation}
It follows from here, in view of (\ref{b56})--(\ref{b59}), that
\begin{equation}
\label{b71}
E_n (z) = n^2 + \sum_{k=1}^6 P_k (z) \frac{1}{n^{2k}} + O(1/n^{14}),
\end{equation}
where
\begin{equation}
\label{b72}
P_1 (z) =-\frac{z^2}{4}, \quad
P_2 (z) = -\frac{2z^2 +3z^4}{32}, \quad
P_3 (z) =-\frac{z^2 +5z^4}{64 },
\end{equation}
$$
P_4 (z) = \frac{-2z^2-21z^4+28z^6}{512},
\quad
P_5 (z) = \frac{-8z^2 -144 z^4 +1920 z^6 + 153 z^8}{8192},
$$
$$
P_6 (z) =
\frac{-2z^2 - 55z^4 + 5192 z^6 +880 z^8}{8192}.
$$
See further discussion in Section 7.3.

\section{Analytic continuation of eigenvalues and Regularized
Trace}

1. Each eigenvalue $E_k (z),$ as we have seen in Proposition \ref{prop1},
is well defined and simple if $|z| \leq R_k = k^{1-\alpha}/8M. $
We are going to show that it is
possible to continue $E_k (z) $ analytically as $z$
is moving along a smooth curve which goes around singular points
$\zeta \in S,$ where S is a countable set without a finite point of
accumulation.

Fix $n \in \mathbb {N} $ and consider the rectangle
$$
W \equiv W_n= \{\lambda \in \mathbb{C}:\; -n < Re \, \lambda < n^2 + n, \;
|Im \, \lambda | < n \}.
$$
By Proposition \ref{prop2}
the projector
$$
P_* (z) = \frac{1}{2\pi i} \int_{\partial W} (\lambda -L-zB)^{-1}
d\lambda
$$
is well defined for $z\in \Delta_n $ and
$$
\dim P_* (z) = n.
$$

Consider the analytic functions
\begin{equation}
\label{37.1}
\sigma_j (z) = Trace \left (
\frac{1}{2\pi i} \int_{\partial W}
\lambda^j (\lambda -L-zB)^{-1} d\lambda \right ), \quad 1 \leq j \leq n.
\end{equation}
If $|z|$ is small, say $|z| < \varepsilon < R_1,$  then
\begin{equation}
\label{37.2}
\sigma_j (z) = \sum_{k=1}^n  (E_k (z))^j, \quad 1 \leq j \leq n,
\end{equation}
where all $E_k (z) $ are well defined. Moreover,
\begin{equation}
\label{37.3}
\prod_1^n (\lambda - E_k) = \sum_{0}^n Q_{n-j} (E) \lambda^j,
\end{equation}
where $ \{Q_i\}_1^n, \; Q_0 \equiv 1, $ are symmetric polynomials
of $ \{E_k\}_1^n.$ But $ \{\sigma_j\}_1^n $ is a basis system of
symmetric polynomials (see, e.g. \cite{MD}), and therefore,
\begin{equation}
\label{37.5}
Q_j = q_j (\sigma )
\end{equation}
are polynomials of $\sigma$'s.
Thus
\begin{equation}
\label{37.6}
\prod_1^n (\lambda - E_k) = \sum_{0}^n q_{n-j} (\sigma(z)) \lambda^j,
\end{equation}
at least for small $|z|,$ say $|z| < \varepsilon.$
However, the coefficients
$c_j (z) = q_{n-j} (\sigma (z)) $ are
well--defined by (\ref{37.5}), (\ref{37.1})
in the entire disk $\Delta_n $ and analytic there.
The factorization (\ref{37.3}) becomes
\begin{equation}
\label{38.1}
\prod_1^n (\lambda - E_k (z)) =
\sum_{0}^n c_j (z) \lambda^j,
\end{equation}
and the equation
\begin{equation}
\label{38.2}
c(z,\lambda): =\sum_{0}^n c_j (z) \lambda^j =0, \quad |z| \leq R_n,
\end{equation}
defines over $\Delta_n $ the surface
\begin{equation}
\label{38.3}
G_n = \{ (\lambda,z) \in \mathbb{C} \times \Delta_n\;: \quad c(z,\lambda)=0
\}
\end{equation}
with $n$ sheets and possible branching points $z_*$
if the polynomial
$\sum_{0}^n c_j (z_*) \lambda^j $
has multiple roots. Such a point $z_* $
is a root of the resultant
\begin{equation}
\label{39.2}
r(z) = R(c(z,\cdot), c^\prime_\lambda (z, \cdot))
\end{equation}
of the polynomial $c(z,\lambda) $
and its derivative $c^\prime_\lambda.$
Notice that $r(z) $ is an analytic function of $z,\; |z|\leq R_n, $
because the resultant is a polynomial of $c_j (z) \in $(\ref{38.2}).
If $z=0,$ then
\begin{equation}
\label{39.3}
\sum_0^n c_j (0) \lambda^j = \prod_1^n (\lambda - k^2),
\end{equation}
and all zeros are simple. Therefore, $ r(0) \neq 0,$
so the resultant $r(z)$ is not identically zero.
Thus the set
\begin{equation}
\label{39.4}
\Sigma_n = \{z\in \Delta_n \; : \;\; r(z) = 0 \}
\end{equation}
is finite.
By Proposition \ref{prop1}
we can conclude that
\begin{equation}
\label{39.5}
\Sigma_n \subset \Sigma_{n+1} \quad \text{and} \quad
\Sigma_{n+1} \cap \Delta_n = \Sigma_n.
\end{equation}
Thus the set
\begin{equation}
\label{40.1}
S= \bigcup \Sigma_n
\end{equation}
is countable and has no finite points of accumulation.

We have proved the following.
\begin{Proposition}
\label{prop10}
Under the conditions of Proposition \ref{prop1},
there is a countable set $S$ without finite accumulation points
such that if
$$\gamma= \{z(t)\; : \; 0 \leq t \leq T\}, \quad  z(0)= 0, \quad
\gamma \cap S = \emptyset $$ is a smooth curve then each
eigenvalue function $E_k (z), \; E_k (0) = k^2, $ can be extended
analytically along the curve $\gamma.$
\end{Proposition}

2. We define {\em Spectral Riemann Surface (SRS)}
of the pair $(L,B)$ as
\begin{equation}
\label{40.4}
G = \{(\lambda,z)\in \mathbb{C}^2 \;: \quad
(L+zB)f = \lambda f,  \quad
 f \in \ell^2 (\mathbb{N}), \; f\neq 0 \}.
\end{equation}

\begin{Proposition}
\label{prop11} Under the conditions of Proposition \ref{prop1},
for each $z \not \in S $ the surface $G$ has infinitely many
sheets over a neighborhood $U_\varepsilon \ni z$ for small enough
$\varepsilon (z) >0. $ Each branching point $z_* \in S $ is of
finite order.
\end{Proposition}

\begin{proof}
Everything has been already explained. The surface $G$ over
$\Delta_n $ is defined by (\ref{38.2}), and, by
(\ref{38.2})--(\ref{39.4}),  $\lambda (z) $ has branching points
$z_* \in \Delta_n $ of order $\leq n.$
\end{proof}

3. We follow the 1975 Sch\"afke construction (see \cite{MSW}, pp. 88--89),
as it is presented by H. Volkmer \cite{V4},
to analyze whether the Spectral Riemann Surface $G$ is {\em irreducible}.

Let $k,j \in \mathbb{N}.$ We call
$k$ and $j$ {\em equivalent}, $ k \sim j,$
if there is a smooth curve
$$
\varphi:\;[0,T] \to \mathbb{C}\setminus S, \quad
\varphi (0) =\varphi (T) =0,
$$
such that the analytic continuation of $E_k (z) $ along $\varphi $
leads to $E_j (z). $ (A Spectral Riemann Surface $G$ is {\em
irreducible} if $\mathbb{N} $ is the only equivalence class, i.e.,
$k \sim j $ for any $k,j \in \mathbb{N}.$)

Such construction, carried for each $k \in \mathbb{N},$
defines a mapping
$$\pi_\varphi: \mathbb{N} \to \mathbb{N}, \quad \pi (k) = j $$
such that $\pi_{\varphi^{-1}} (j) = k, $
where $\varphi^{-1} (t) = \varphi (T-t).$
With $R_n \to \infty, $ we have for some $n$ that
$\max_{[0,T]} |\varphi (t) | \leq R_n. $
Therefore, by Proposition \ref{prop1},
\begin{equation}
\label{43.1}
\pi_{\varphi} (k) = k \quad \text{if} \quad k >n.
\end{equation}

\begin{Lemma}
\label{lem11}
Let $\mathcal{M}$ be an equivalence class
(or union of equivalence classes),
and $ n \in \mathbb{N}.$
Then the function
\begin{equation}
\label{43.2}
\widetilde{E}_n (z) = \sum_{k \in \mathcal{M}, k\leq n} E_k (z)
\end{equation}
(which is well--defined and analytic for small enough $|z|$)
can be extended analytically on the disk
$\{z: |z| \leq R_n\}, \; R_n = n^{1-\alpha}/(8M).$
\end{Lemma}

\begin{proof}
Take any smooth curve
$\varphi:[0,T] \to \Delta_n \setminus S, $
such that $\varphi (0)= \varphi (T) = 0. $
Then $ \pi = \pi_{\varphi} : \mathcal{M} \to \mathcal{M} $
is a bijection, and (\ref{43.1}) holds,
so $\pi $ permutes the finite set $\{k \in \mathcal{M}: k \leq n\}.$
Therefore, $\widetilde{E}_n (z) $ can be continued analytically,
term by term in (\ref{43.2}), and the result will be
$$ \sum_{k \in \mathcal{M}, k \leq  n} E_{\pi(k)} (z) =
\sum_{j \in \mathcal{M}, j \leq  n} E_j (z) =\widetilde{E}_n (z), \quad z \in \Delta_n \setminus S,
$$
i.e., the same function. By Proposition \ref{prop2}, if
$|z| \leq R_n, $ then we have exactly $n$ eigenvalues
on the left of the line $h_n = \{ Re\, \lambda = n^2 +n \},$
and all of them
lie in the rectangle $ W_n. $ Therefore,
\begin{equation}
\label{44.2}
\left | \widetilde{E}_n (z) \right |  \leq n(n^2 +2n).
\end{equation}
So the function $ \widetilde{E}_n (z) $
is analytic and bounded on $\Delta_n \setminus S, $
while the set $\Delta_n \cap S $ is finite.
Thus, it is analytic in the disk $\Delta_n. $
\end{proof}

The inequality (\ref{44.2}) cannot be improved essentially because
$$
\sum_1^n E_k (0) = \sum_1^n k^2 = n(n+1)(2n+1)/6 \sim n^3.
$$
However, we can regularize $\tilde{E}_k (z) $ by considering
$\tilde{E}_k (z) - \tilde{E}_k (0),$
where $\tilde{E}_k (0) $ is real.

Again by Proposition \ref{prop2},
if $|z| \leq R_n, $ then the operator $L+zB $
has $n$ eigenvalues that lie in the rectangle $W_n, $ so
the absolute value of the imaginary part of each of these eigenvalues
is less than $n.$
Therefore,
\begin{equation}
\label{44.3}
\left | Im \left
(\widetilde{E}_n (z)- \widetilde{E}_n (0) \right ) \right | \leq n^2.
\end{equation}

By Borel--Caratheodory theorem (see Titchmarsh \cite{Titch},
Ch.5, 5.5 and 5.51),
if $g(z) $ is analytic in the disk $|z| < R, $
$g(0) =0$ and $|Im \,g(z)| \leq C,  $
then $|g(z)| \leq 2C  $
for $|z| \leq R/2. $
Thus (\ref{44.3}) implies
\begin{equation}
\label{44.5}
\left |
\widetilde{E}_n (z)- \widetilde{E}_n (0) \right | \leq 2n^2,
\quad \text{for} \quad |z| \leq R_n/2.
\end{equation}

This conclusion is valid for each equivalence class,
or union of equivalence classes $\mathcal{M};$
in particular, for $\mathcal{M}= \mathbb{N}.$ \vspace{5mm}

4. {\it Definition of the regularized trace $tr (z).$} Now we are
ready to define an entire function $tr (z), $ the regularized
trace of $L+zB, $ under the conditions (\ref{10.1}) and
(\ref{10.2}) with $\alpha <1/2,$ or (\ref{6.1}) with $ \alpha =
1/2.$

For small $z,\; |z| \leq R_1 = 1/(8M),$ all $E_n (z) $ are well
defined, and
\begin{equation}
\label{60.1}
tr (z) = \sum_{n=1}^\infty \left (E_n (z) - n^2 \right )
=\sum_{n=1}^\infty \left ( \sum_{k=1}^\infty a_{2k} (n) z^{2k} \right )
\end{equation}
$$ =\sum_{n=1}^\infty  \left ( a_2 (n) z^2 +  a_4 (n) z^4 +
\sum_{k=3}^\infty a_{2k} (n) z^{2k} \right )  =
$$
$$
= z^2 \cdot \lim_{p\to \infty}
\varphi_2 (p) + z^4 \cdot \lim_{p\to \infty} \varphi_4 (p)   +
\sum_{n=1}^\infty  \sum_{3}^\infty \cdots
$$
By (\ref{p31a})--(\ref{p33b}), the latter limits are well defined
and the third term is an absolutely convergent series. Indeed, by
(\ref{45.1}),  Lemma \ref{lem12}, we have for $|z| < 1/ (8M) $
that
$$
\sum_{n=1}^\infty \left ( \sum_{k=3}^\infty |a_{2k} (n)| \right )
|z|^{2k} \leq \sum_{n=1}^\infty  \left ( \sum_{k=3}^\infty
\frac{(8M)^{2k}} {n^{2k(1-\alpha)-1}} \right ) \frac{1}{(8M)^{2k}}
$$
$$
= \sum_{n=1}^\infty  \frac{n}{n^{6(1-\alpha)}} \cdot \frac{1}{1-n^{\alpha -1}}
< \infty  \quad \text{for} \quad \alpha < 2/3.
$$
Therefore (\ref{60.1}) defines $tr (z) $
as an analytic function in the disk
$ |z| \leq 1/(8M). $

Fix $ N \in \mathbb{N} $ and consider
the analytic function
$\tilde{E}_N (z), \;z \in \Delta_N, $
given by Lemma \ref{lem11}
in the case where $ \mathcal{M} = \mathbb{N}. $
For small $|z|$ we have
\begin{equation}
\label{60.2}
tr (z) = \widetilde{E}_N (z) -\widetilde{E}_N (0) +
\sum_{n=N+1}^\infty \left (E_n (z) - n^2 \right ).
\end{equation}
The same formula gives the analytic extension of $tr (z) $
on $\Delta_N $ because the series on the right side of (\ref{60.2})
converges uniformly on $\Delta_N. $
Indeed, with $E_n (z) = a_2 (n) z^2 + a_4 (n) z^4 + \cdots, $
we have
\begin{equation}
\label{60.3}
\sum_{N+1}^\infty \left (E_n (z) - n^2 \right ) =
\left (\sum_{N+1}^\infty a_2 (n) \right ) z^2 +
\left (\sum_{N+1}^\infty a_4 (n) \right ) z^4 +
\sum_{n=N+1}^\infty  \sum_{k=3}^\infty a_{2k} (n) z^{2k}.
\end{equation}
By (\ref{p8}) we obtain, for $n \geq N+1 $ and
$|z| \leq R_N = N^{1-\alpha}/(8M),$ that
\begin{equation}
\label{60.4} \sum_{k=3}^\infty |a_{2k} (n) |\, |z|^{2k} \leq
\sum_{k=6}^\infty \frac{4k+2}{n^{(1-\alpha)k-1}} (4M)^k R_N^k \leq
C(N, \alpha)  \left (\frac{1}{n} \right )^{6(1-\alpha)-1},
\end{equation}
where $C(N, \alpha) = N^{6(1-\alpha)} \sum_{k\geq 6} (4k+2)2^{-k}
<\infty.$ Now, in view of (\ref{60.3}), the estimate (\ref{60.4})
implies that the series in (\ref{60.2}) converges uniformly in
$\Delta_N $ if $\alpha \in [0,1/2],$ thus $tr(z) $ can be extended
analytically in the disk $\Delta_N. $ Since $\cup_N \Delta_N =
\mathbb{C} $ {\em this defines $tr (z) $ as an entire function.}
\vspace{4mm}

5. {\em Proof of Theorem \ref{thm1}.}
According to the previous subsection, $tr (z) $
is an entire function. Therefore, it is enough to prove
(\ref{7.2}) 
only for small $|z|, $
or to evaluate its Taylor coefficients.
By (\ref{60.3})
$
tr (z) =  \sum_1^\infty A_{2k} z^{2k} $ where
$$
A_{2k} = \sum_{n=1}^\infty a_{2k} (n) = \lim_{p \to \infty}
\varphi_{2k} (p). $$
If $ \alpha < 1/2,$
then we have,
by (\ref{p32b}) and (\ref{p33b}), that
$$
\lim_{p\to \infty} \varphi_{2k} (p) = 0 \quad k=1,2, \ldots,
$$
and therefore, $ tr (z) \equiv 0. $

If $\alpha = 1/2, $ then (\ref{p32b}) and (\ref{p33b}) imply 
$$
\lim_{p\to \infty} \varphi_{2k} (p) = 0 \quad k=2, \ldots,
$$
and by (\ref{p31a}), if the limit $\ell = \lim b_k c_k/k $ exists, then
$$
\lim_{p\to \infty} \varphi_{2} (p) =
\lim_{p\to \infty} \left ( -\frac{b_p c_p}{2p+1} \right )
= - \frac{\ell}{2},
$$
so $ tr(z) = -(\ell/2) z^2.$
This completes the proof of Theorem \ref{thm1}.

\section{Spectral Riemann Surfaces}
In our analysis of the regularized trace it was important to see
that by
inequality (\ref{45.1}) from Lemma \ref{lem12}
$$
a_k (n) = \frac{1}{k!} \left | E_n^{(k)} (0) \right |
\leq (8M)^k n^{1-k(1-\alpha)},
$$
so the series
\begin{equation}
\label{45.4}
\sum_{n=1}^\infty \left | E_n^{(k)} (0) \right |  < \infty
\quad \text{if} \quad \alpha < 1- 2/k.
\end{equation}
Therefore,
 for every subset $\mathcal{M} \subset \mathbb{N}, $
the partial sum
\begin{equation}
\label{45.5}
\mathcal{E}^{(k)} (\mathcal{M}) =
\sum_{m\in \mathcal{M}} E^{(k)}_m (0)
\end{equation}
is well defined.

On the other hand, (\ref{44.5}) and the Cauchy inequality for the
Taylor coefficients yield
\begin{equation}
\label{45.6}
\frac{1}{k!} \left | \widetilde{E}_n^{(k)} (0) \right |
\leq \frac{2n^2}{(R_n/2)^k} = 2(16M)^k n^{2-k(1-\alpha)}.
\end{equation}
So, if $\alpha < 1-2/k, $ then
\begin{equation}
\label{45.7}
\lim_n \left | \widetilde{E}_n^{(k)} (0) \right | =
\lim_n \left |
\sum_{m\in \mathcal{M}, m\leq n} E^{(k)}_m (0) \right | = 0.
\end{equation}
Therefore, the following statement is true.

\begin{Proposition}
\label{prop14}
If $ \alpha < 1 - 2/k , $
then we have for each equivalence class $\mathcal{M}$
of the Spectral Riemann Surface of the pair $(L,B) $ that
\begin{equation}
\label{45.8}
\mathcal{E}^{(k)} (\mathcal{M}) \equiv
\sum_{m\in \mathcal{M}} E^{(k)}_m (0) =0.
\end{equation}
\end{Proposition}

4. Finally we show
that some Spectral Riemann Surfaces are irreducible, which is the
claim of Theorem \ref{thm3}.

{\em Proof of Theorem \ref{thm3}.}
First we consider the case where
(\ref{6.1}) holds with $ \alpha = 1/2.$
By Proposition \ref{prop14},
Theorem \ref{thm3} will be proved
if we show that there is no proper subset
$ \mathcal{M} \subset \mathbb{N} $
with the property (\ref{45.8})
for a fixed $k >2/(1-\alpha).$
Indeed, then $\mathbb{N}$ will be the
only one equivalence class, which implies that
the Spectral Riemann Surface is irreducible.

If $ \alpha = 1/2 $ then $k=6 $ is the least even $k$
for which $k >2/(1-\alpha).$
By (\ref{b58}) we have
\begin{equation}
\label{200}
\frac{1}{k!} E_n^{(6)}(0) = a_6 (1/2,n)= \psi (n)- \psi (n-1),
\qquad  n=2,3, \ldots,
\end{equation}
where
\begin{equation}
\label{202}
\psi (n) = - \frac{1}{(2n-1)(2n+1)^5 (2n+3)}.
\end{equation}
On the other hand,
from (\ref{25b}), with $\alpha = 1/2,$
it follows that
\begin{equation}
\label{203}
a_6 (1/2,1) = \psi (1) = -\frac{1}{5 \cdot 3^5}.
\end{equation}

In view of (\ref{200})--(\ref{203}),
\begin{equation}
\label{204}
a_6 (1/2,1) < 0, \quad a_6 (1/2,n) >0 \quad \text{for} \quad n \geq 2,
\end{equation}
and
$$
\sum_{n=2}^\infty  a_6 (1/2,n) = - a_6 (1/2,1).
$$
Certainly,
\begin{equation}
\label{205}
\text{if} \quad \sum_{n\in \mathcal{M}} a_6 (n) = 0
\quad \text{then}\quad \mathcal{M} = \mathbb{N}.
\end{equation}
This proves Theorem \ref{thm3} for $ \alpha = 1/2.$

If $\alpha \in [0,1/2),$ then $4 >2/(1-\alpha),$
so, in view of Proposition \ref{prop14}
and the above discussion,
the Spectral Riemann Surface corresponding to  $\alpha \in [0,1/2)$
will be irreducible if all but one terms of the sequence
$$ a_4 (\alpha,n)= \frac{1}{4!} E_n^{(4)} (0) $$
have the same sign. Below, in Lemma~\ref{lem15},
we show that this is true if $\alpha \in [0,0.085]$
and $\alpha \in [(2-\sqrt{2})/4, 1/2],$
which completes the proof of Theorem \ref{thm3}.
\vspace{3mm}

5. For convenience we set $\gamma = 2 \alpha $ and
$$
\tilde{a}_4 (\gamma,n) = a_4 (\gamma/2,n), \quad \tilde{\varphi}_4
(\gamma,n) = \varphi_4 (\gamma/2,n).
$$
\begin{Lemma}
\label{lem15}
Under the above notations we have
\begin{equation}
\label{101}
\tilde{a}_4 (\gamma,1) =\tilde{\varphi}_4 (\gamma,1)
=\frac{1}{27} - \frac{2^\gamma}{72} >0, \quad \gamma \in [0,1];
\end{equation}
\begin{equation}
\label{102}
\tilde{a}_4 (\gamma,2) <0, \quad \gamma \in [0,1];
\end{equation}
\begin{equation}
\label{103}
\tilde{a}_4 (\gamma,n) >0 \quad \text{if} \quad \gamma \in [0,0.1717], \;\;
n\geq 3;
\end{equation}
\begin{equation}
\label{104}
\tilde{a}_4 (\gamma,n) <0 \quad \text{if} \quad \gamma \in
[(\sqrt{2}-1)/\sqrt{2},1], \;\; n \geq 3.
\end{equation}

\end{Lemma}

\begin{proof}

By (\ref{p27}) and (\ref{p32}) we have that (\ref{101}) holds,
and moreover,
\begin{equation}
\label{105}
\tilde{a}_4 (\gamma,n) = \tilde{\varphi}_4 (\gamma,n)-
\tilde{\varphi}_4 (\gamma,n-1),
\end{equation}
where
\begin{equation}
\label{106}
\tilde{\varphi}_4 (n) = \frac{n^{2\gamma}}{(2n+1)^3} -
\frac{n^{\gamma}(n+1)^{\gamma}}{(2n+1)^2(4n+4)}-
\frac{(n-1)^{\gamma}n^{\gamma}}{4n(2n+1)^2}.
\end{equation}
In particular,
$$
\tilde{a}_4 (\gamma, 2) =\tilde{\varphi}_4 (\gamma,2)-
\tilde{\varphi}_4 (\gamma,1)=
\left ( \frac{2^{2\gamma}}{5^3} - \frac{6^{\gamma}}{300} - \frac{2^\gamma}{200}
\right ) - \left  (\frac{1}{27} - \frac{2^\gamma}{72} \right ).
$$
Graphing $\tilde{a}_4 (\gamma, 2)$ one can easily see that
(\ref{102}) holds.
In the same way one can verify that the following inequalities hold:
\begin{equation}
\label{103a}
\tilde{a}_4 (\gamma,m) >0 \quad \text{if} \quad \gamma \in [0,0.1717], \;\;
m=3,4,5,6,
\end{equation}
and
\begin{equation}
\label{104a}
\tilde{a}_4 (\gamma,m) <0 \quad \text{if} \quad \gamma \in
[(\sqrt{2}-1)/\sqrt{2},1], \;\; m=3,4,5,6.
\end{equation}

In order to prove (\ref{103}) and (\ref{104}) for each $n > 6 $
we study
the sign of partial derivative $\partial \varphi_4/\partial n.$
Set
\begin{equation}
\label{107}
b(\gamma,n) = n^{2-2\gamma} (2n+1)^2 \cdot \frac{\partial \tilde{\varphi}_4}
{\partial n} (\gamma, n);
\end{equation}
then
\begin{equation}
\label{108}
b(\gamma,n) = - \frac{3}{2} \left ( 1+ \frac{1}{2n} \right )^{-2} +
\gamma \left ( 1+ \frac{1}{2n} \right )^{-1} -
\end{equation}
$$
\frac{\gamma}{4} \left ( 1+ \frac{1}{n} \right )^{\gamma-1} -
\frac{c-1}{4} \left ( 1+ \frac{1}{n} \right )^{\gamma-2} +
\frac{1}{2} \left ( 1 + \frac{1}{n} \right )^{\gamma-1}
\left ( 1+ \frac{1}{2n} \right )^{-1} -
$$
$$
\frac{\gamma}{4} \left ( 1- \frac{1}{n} \right )^{\gamma-1} +
\frac{1}{2} \left ( 1- \frac{1}{n} \right )^{\gamma}
\left ( 1+ \frac{1}{2n} \right )^{-1} -
\frac{\gamma-1}{4} \left ( 1 - \frac{1}{n} \right )^{\gamma}.
$$
The power series expansion of $b(\gamma, n) $ about $n= \infty $ is
\begin{equation}
\label{109}
b(\gamma,n) =
\sum_{k=2}^\infty b_k (\gamma) (1/n)^k,
\end{equation}
where
\begin{equation}
\label{111}
b_2(\gamma) = \frac{5-22\gamma +18 \gamma^2 - 4\gamma^3}{8},
\end{equation}
\begin{equation}
\label{112}
b_3(\gamma) =
\frac{-10+25 \gamma - 14 \gamma^2 + 2 \gamma^3}{8}.
\end{equation}

By (\ref{108}), estimating from above $|b_k (\gamma)|, $
we obtain
\begin{equation}
\label{114}
b_k(\gamma) \leq \frac{3}{2}\cdot \frac{k+1}{2^k} +
\frac{\gamma}{2^k} + \frac{\gamma}{4} +
\frac{1-\gamma}{4} (k+1) +
\end{equation}
$$
\frac{1}{2}
\left ( \frac{1}{2^k} + 2 \gamma \right )+  \frac{\gamma}{4} +
\frac{1}{2}\left ( \frac{1}{2^k} + \gamma \right ) +
\frac{1-\gamma}{4} \cdot \frac{\gamma}{k} ,
$$
where each term comes from the expansion of the corresponding term in
(\ref{108}).

For example, consider
\begin{equation}
\label{115}
\left (1-\frac{1}{n} \right )^\gamma \left (1+\frac{1}{2n} \right )^{-1}=
\left [1 + \sum_{i=1}^\infty   \binom{\gamma}{i}
\left ( \frac{1}{n} \right )^i  \right ]
\sum_{j=0}^\infty  2^{-j}
\left (- \frac{1}{n} \right )^j.
\end{equation}
Since $ 0 \leq \gamma \leq 1 $ we have
\begin{equation}
\label{116}
\left |\binom{\gamma}{i} \right | = \frac{\gamma}{i} \cdot
\frac{|\gamma -1|}{1} \cdot \frac{|\gamma -2|}{2} \cdots
\frac{|\gamma -(i-1)|}{i-1} \leq \frac{\gamma}{i}.
\end{equation}
Thus the absolute value of the coefficient of $(1/n)^k $
in (\ref{115}) does not exceed
$$
\frac{1}{2^k} + \frac{\gamma}{1} \cdot \frac{1}{2^{k-1}}+
\frac{\gamma}{2} \cdot \frac{1}{2^{k-2}} +
\frac{\gamma}{3} \cdot \frac{1}{2^{k-3}} + \cdots + \frac{\gamma}{k} \leq
$$
$$
\frac{1}{2^k} + \frac{\gamma}{2}
\left ( \frac{1}{2^{k-2}} +\frac{1}{2^{k-2}} + \frac{1}{2^{k-3}} +
\cdots + 1 \right ) =
\frac{1}{2^k} + \gamma.
$$

The inequality (\ref{114}) may be written as
\begin{equation}
\label{118}
b_k(\gamma) \leq \frac{3}{2}\cdot \frac{k+1}{2^k} +
\frac{1+\gamma}{2^k} + 2\gamma +
\frac{1-\gamma}{4} (k+1) +
\frac{\gamma}{4k}.
\end{equation}
Since
$$
\sum_{k=4}^\infty x^k = \frac{x^4}{1-x}, \qquad
\sum_{k=4}^\infty (k+1) x^k = \left ( \frac{x^5}{1-x} \right )^\prime
=\frac{5x^4-4x^5}{(1-x)^2},
$$
we obtain, by (\ref{118}), that
\begin{equation}
\label{122}
\sum_{k=4}^\infty |b_k (\gamma)| n^{-k}  \leq M (\gamma, n)
\cdot \frac{1}{n^3},
\end{equation}
where
\begin{equation}
\label{123}
M (\gamma, n) =
\frac{3(10n-4)}{16(2n-1)^2} + \frac{1+\gamma}{8(2n-1)} +
\frac{33\gamma}{16(n-1)} + \frac{1-\gamma}{4} \cdot \frac{5n-4}{(n-1)^2}.
\end{equation}
Thus we have
\begin{equation}
\label{124}
nb_2 (\gamma) + b_3 (\gamma)- M (\gamma, n) \leq n^3
b(\gamma,n) \leq
nb_2 (\gamma) + b_3 (\gamma) + M (\gamma, n).
\end{equation}

On the other hand,
$$
8b_2 (\gamma) = (5-2\gamma)(\gamma - (1-1/\sqrt{2}))
(\gamma - (1 + 1/\sqrt{2})),
$$
and therefore,
\begin{equation}
\label{125}
b_2 (\gamma) >0 \quad \text{for} \;\; \gamma \in [0,1-1/\sqrt{2}),
\quad
b_2 (\gamma) <0 \quad \text{for} \;\; \gamma \in (1-1/\sqrt{2},1].
\end{equation}
One can easily see, for each fixed $\gamma \in [0,1],$ that
$M(\gamma,n)$ is a decreasing function of $n.$
This fact leads,
in view of (\ref{124}) and (\ref{125}),
to the following inequalities:
\begin{equation}
\label{126}
0 <  6b_2 (\gamma) + b_3 (\gamma)- M (\gamma, 6) \leq n^3
b(\gamma,n),
\qquad \gamma \in [0,0.19), \;\;n\geq 6
\end{equation}
and
\begin{equation}
\label{127}
n^3 b(\gamma,n) \leq
6b_2 (\gamma) + b_3 (\gamma) + M (\gamma, 6) < 0, \qquad
\gamma \in (1-1/\sqrt{2},1], \;\; n\geq 6.
\end{equation}
(We checked the left inequality in (\ref{126}) and
the right inequality in (\ref{127}) numerically by graphing the
corresponding functions of $\gamma.$)

In view of (\ref{107}) and (\ref{126}),
$\partial \tilde{\varphi}/\partial n (\gamma, n ) >0  $
if $\gamma \in [0, 0.19] $ and $ n \geq 6, $
so $\tilde{\varphi} (\gamma,n ) $ increases with $n.$
Therefore, for each  $\gamma \in [0, 0.19] $  and $ n > 6,$
we obtain by (\ref{103a}) that
$\tilde{a} (\gamma, n) > \tilde{a} (\gamma, 6) >0, $
which proves (\ref{103}).

In a similar way (\ref{127}) implies that
$ \tilde{\varphi} (\gamma,n ) $ decreases with $n$  if
$\gamma \in [1-1/\sqrt{2},1]$ and $n\geq 6. $
Thus, in view of (\ref{104a}), we obtain that
$\tilde{a} (\gamma, n) < \tilde{a} (\gamma, 6) < 0, $
for
$\gamma \in [1-1/\sqrt{2},1]$ and $n\geq 6, $
which proves (\ref{104}).
This completes the proof of Lemma \ref{lem15}.

\end{proof}

\section{Conclusion; comments and questions}

1. So far in our analysis we focused on the tri--diagonal matrices
given by (\ref{10.1}) and (\ref{10.2}), or (\ref{6.1}) with
$\alpha < 1,$ or even with $ \alpha \leq 1/2.$
The Whittaker--Hill matrices (\ref{7.1}) satisfy
(\ref{10.1}) and (\ref{10.2}) with $\alpha =1, \; M= 4+t.$
Proposition \ref{prop1} tells us that the eigenvalues
$E_n (z), \; n\in \mathbb{N}, $ are analytic functions
in the disk $\Delta = \{|z| <1/(8M) \}, $
and nothing more. But these matrices
come from the differential operator
\begin{equation}
\label{30.2}
Ay = - y^{\prime \prime} + q(x) y,
\end{equation}
considered with
\begin{equation}
\label{30.3}
q(x) = a \cos 2x + b \cos 4x, \qquad a= - 4zt, \;\; b = - 2z^2.
\end{equation}

Let $q(x) $ be a real analytic periodic function of period $\pi.$
Of course, then $q$ extends analytically in a
neighborhood of $I= [0,\pi], $ say, in
\begin{equation}
\label{31.1}
G_\varepsilon = \{w= x+iy \;: \quad
-\varepsilon \leq x \leq \pi + \varepsilon, \quad
-\varepsilon \leq y \leq \varepsilon \}, \quad
\exists \varepsilon >0.
\end{equation}
In other words, $q$ is in the Banach space
$A (G_\varepsilon) $ of
all functions that are continuous in
$ G_\varepsilon $ and analytic in its interior,
with the norm
$$
\|f\| = \max \{ |f(w)|: \; w \in  G_\varepsilon \}.$$

Consider the boundary conditions
$$ Per^+: \quad y(0) = y(\pi), \;\; y^\prime (0) = y^\prime (\pi),$$
$$ Per^-: \quad y(0) = -y(\pi), \;\; y^\prime (0) =- y^\prime (\pi),$$
$$ Dir: \quad y(0) = y(\pi) = 0. $$

To be certain, let us talk only about the periodic boundary
conditions $Per^+ ,$ and let us consider the (invariant) subspace
of even functions. Then
the operator  (\ref{30.2})
has eigenvalue functions $E_n (z), \; E_n (0) = (2n)^2, \; n=1,2, \ldots.$

H. Volkmer \cite{V2}
proved that if $q $ is a real analytic function, then $E_n (z) $
is well defined as an analytic function in
the disk
$$ \Delta_n = \{z: \; |z| \leq R_n\}, \quad R_n = a n^2,\quad a>0.
$$
Careful analysis of the proof in \cite{V2} shows that a stronger
quantitative statement holds.

\begin{Proposition}
\label{prop17}
If
\begin{equation}
\label{31.1a}
 q \in A (G_\varepsilon), \quad \varepsilon > 0,
\end{equation}
then the eigenvalues  $E_n (q) $ of the operator
(\ref{30.2})
are well defined
if $q$ is real--valued on $[0,\pi]$
and small by norm. Moreover, for each $n,$
$E_n (q) $ can be extended as an analytic function of $q$
in the ball
$$ B(R_n) = \{q \in A(G_\varepsilon): \;\|q\|\leq R_n \},
$$
with $ R_n = a n^2, \;\;  a= a(\varepsilon ) >0. $
\end{Proposition}

As soon as we have this Proposition,
we can consider the potentials (\ref{30.3})
as elements of $A(G_\varepsilon),$
with, say, $\varepsilon = 1/4. $
Then
\begin{equation}
\label{32.4}
\|4zt \cos 2x + 2 z^2 \cos 4x \| \leq
4|zt| e^{2\cdot\frac{1}{4}} + 2 |z|^2 e^{4\cdot\frac{1}{4}}
\leq  7 \left ( |zt|+ |z|^2 \right),
\end{equation}
and therefore, if
\begin{equation}
\label{33.1}
 |tz|+ |z|^2  \leq \frac{a}{7}\, n^2,
\end{equation}
then
\begin{equation}
\label{33.2}
e_n (z) = E_n (q), \quad  q =4zt \cos 2x + 2z^2 \cos 4x,
\end{equation}
is an analytic function of $z.$
Choose
\begin{equation}
\label{33.3}
        R_n = n(1+4|t|/a)^{-1};
\end{equation}
then
\begin{equation}
\label{33.4}
z \in \Delta_n = \{z: |z| \leq R_n \}  \quad \Rightarrow \;\; (\ref{33.1}),
\end{equation}
and therefore,
the function $e_n (z) $ is analytic in the disk $\Delta_n.$

We explained the following statement
(which is stronger than its analogue coming from
Proposition \ref{prop1}).

\begin{Proposition}
\label{prop18}
Under the conditions (\ref{7.1}) the spectrum of the operator
(\ref{1}) is discrete.
The function $e_n (z)\in (\ref{33.2}) $ is analytic in
$\Delta_n \in (\ref{33.4}) ,$
and
$$
e_n (0) = n^2, \quad |e_n (z) - n^2 | \leq n
\quad \text{if} \;\; z \in \Delta_n.
$$
\end{Proposition}
\vspace{3mm}

2. Of course, the claim of Proposition \ref{prop18}, with $R_n =
an/(a+4|t|),$ is stronger than Proposition \ref{prop1} with $R_n =
1/8. $ This example, together with Remark \ref{rem1}, supports our
belief that, for matrices $(L,B) \in (\ref{10.1}),$ Proposition
\ref{prop1} can be significantly improved, so that to give
analyticity of $E_n (z) \in (\ref{10.3}), (\ref{10.5})$ in the
disk $\Delta_n \in (\ref{10.4})$ with
\begin{equation}
\label{34.3}
R_n = bn^{2-\alpha}, \quad \exists b= b(\alpha)>0.
\end{equation}
If $\alpha = 0 $ this is true, but again it comes from H.
Volkmer's result \cite{V1,V2} for the Mathieu differential
operator which is unitary equivalent to the matrices (\ref{6.1}),
(\ref{6.2}).

However, even in this case, no approach to the proof of
this statement is known in the framework of matrix analysis.
\vspace{3mm}

3. Of course, if the Taylor expansion
$$
E_n (z) = n^2 + \sum_{k=1}^\infty a_{2k} (n) z^{2k}
$$
is known, then one may find the radius of convergence of
$E_n (z) $  as
$$
r_n = \left ( \limsup_k |a_{2k}(n)|^{1/2k} \right )^{-1} .
$$
Proposition \ref{prop3} gives that
$$
|a_{2k}(n)| \leq 8kn\cdot \left ( \frac{8M}{n^{1-\alpha}} \right )^{2k}
$$
for  $(L,B) \in (\ref{10.1}) +(\ref{10.2}).$ However,
if $B \in (\ref{6.1}), $
i.e.,
$$ b_k = c_k = k^\alpha, \qquad   0 \leq  \alpha < 2,
$$
we believe that
\begin{equation}
\label{35.5}
|a_{2k}(n)| \leq n^\gamma
\left ( \frac{A}{n^{2-\alpha}} \right )^{2k},
\quad n=1,2,\ldots, \;\; \gamma > 0, \;A>0,
\end{equation}
Of course, (\ref{35.5}) would imply (\ref{34.3}).\vspace{3mm}

4. Maybe, the representation (\ref{p5}), (\ref{p27}), (\ref{p28})
of Propositions \ref{prop3} and \ref{prop4} could be used in
an attempt to get (\ref{35.5}).
But let us make a couple of elementary remarks
to Propositions \ref{prop3} and \ref{prop4}. \vspace{3mm}

\begin{Remark}
\label{rem2}
{\em It was observed in (\ref{p7}), on the basis of
the representation (\ref{p5}) and (\ref{p27}), (\ref{p28}),
that $a_k (n) \equiv 0 $ for odd $k.$}
This follows also from
the equality
\begin{equation}
\label{36.1}
Sp (L+ zB) = Sp (L-zB), \quad z \in \mathbb{C},
\end{equation}
because (\ref{36.1}) implies
that all $E_n (z) $ are even functions.
\end{Remark}

(In particular, this implies that in formulas like (\ref{b71}) and
(\ref{b72}) the coefficients $P_k(z)$ should be even functions. In
[5], however, formula (8) in Theorem 2.1 has $P_1(z) = (z^3 -
4z)/16,$ so one can conclude that this is not correct even without
knowing the correct formula.)

To get (\ref{36.1}),
consider the unitary operator $U$ defined by
\begin{equation}
\label{36.2}
U e_j = (-1)^j e_j, \quad 1 \leq j < \infty, \quad U^2 = 1.
\end{equation}
Then for each matrix $A = [A(i,j)] $ the operator
$\widetilde{A}= U^{-1} A U = UAU $ has a matrix
$\widetilde{A} (i,j) = (-1)^{i-j} A(i,j). $
In particular,
$ U^{-1} (L+zB) U = L-zB,$
i.e., the operators $L+zB $ and $L-zB $ are similar,
and therefore, (\ref{36.1}) holds.
Of course, this implies that $E_n (z) $ are even functions.
\vspace{3mm}

\begin{Remark}
\label{rem3}
{\em By Propositions \ref{prop3} and \ref{prop4},
the integrals that appear in  (\ref{p7}) and (\ref{p28}) vanish
if $|j-n| > k.$ But}  they vanish even if $ |j-n|> k/2.$
\end{Remark}

After Remark \ref{rem2} we can talk only about even $k,$
say $k=2m. $
Let us focus on (\ref{p28}), i.e., on the integrals
\begin{equation}
\label{37.1a}
I(n;j,k) =
\int_{h_n}
\lambda
\langle R^0_\lambda (BR^0_\lambda)^k e_j, e_j \rangle
d \lambda,
\end{equation}
where $ h_n = \{\lambda \in \mathbb{C}: \; \lambda = n^2 +n +it, \;
t \in \mathbb{R} \}.$

The integrand in (\ref{37.1a}) is a linear combination of
rational functions like (\ref{p29a}) with coefficients depending
on $B,$ where each rational function corresponds to a walk
$ (j_0, j_1, \ldots, j_k) $ from $j$ to $j$
on the
integer grid $\mathbb{Z},$ with steps $\pm 1.$
Indeed, when the operator $R^0(BR^0)^k $ acts on $e_j, $
then (since $R^0 e_\nu = (1/(\lambda-\nu^2))e_\nu $ while
$Be_\nu$ is a linear combination of $e_{\nu -1} $ and
$e_{\nu+1} $)
we get a linear combination of $2^k $ vectors,
each of them coming from some walk
$ (j_0, j_1, \ldots, j_k) $ as
$e_{j_0}\to e_{j_1} \to \cdots \to e_{j_k}.$
Since
$\langle  e_{j_k}, e_j \rangle \neq 0 $
only for $j_k = j,$
we consider further only walks from $j$ to $j.$

Moreover, the argument used to prove the point (iii) in the proof of
Proposition \ref{prop4} shows that
the rational function  $Q$ of (\ref{p29a})
yields a non-zero integral over the line $h_n $ only if
it has poles both on the left and on the right of $h_n, $
and its poles $ j_\nu^2 $
come from
the vertexes of the corresponding walk
$ (j_0, j_1, \ldots, j_k). $
In other words,  if $j<n $ (respectively $j > n$) then
the corresponding walk $j_0=j, j_1, \ldots, j_k=j$
should pass through $n+1 $ (respectively $n$).

Take now any $j$ such that $|j-n|>k/2 .$
If $j < n $ (respectively $j > n$),
then there is no $k$-step walk
from $j$ to $j$ passing through $n+1$
(respectively $n $) because the steps are equal to $\pm 1.$
Thus each of the integrals (\ref{37.1a}) vanishes if $|j-k|> k/2.$
\vspace{3mm}

5. We consider $\alpha \geq 0 $ in (\ref{6.1}) and elsewhere
to have unbounded or non-compact operators $B.$
Of course, Theorems \ref{thm1} and \ref{thm2}
remain valid for $\alpha <0 $ as well.
But then a simpler proof
can be given because for $\alpha <0 $
the restriction $\alpha < 1- 2/k $
holds with $k=2. $
In particular, by (\ref{45.8}), i.e., by Proposition \ref{prop14},
we have
$$
\mathcal{E}^{(2)} (\mathcal{M}) = \sum_{m \in \mathcal{M}}
E_m^{(2)} (0) = 0
$$
for any equivalence class of the Spectral Riemann Surface
of the pair $(L,B) \in (\ref{10.1})+ (\ref{10.2}), \;
\alpha <0. $

Of course, it is easier to study the sign of
$a_2 (\alpha, n) $
than the sign of $a_4 (\alpha, n) $
(compare to Lemma~\ref{lem31}).
By (\ref{p31a}) we have that
$$
a_2 (1) = - \frac{b_1 c_1}{3}, \qquad
a_2 (n) =\frac{b_{n-1} c_{n-1}}{2n-1} - \frac{b_n c_n}{2n+1}.
$$
If $(b_n )$ and $(c_n ) $ are decreasing sequences of positive
numbers, then
$$
a_2 (1) <0, \qquad a_2 (n) > 0, \;\;n \geq 2,
$$
and we can use the same argument as before (see the proof of
Theorem \ref{thm3}) to conclude that the corresponding Spectral
Riemann Surface is irreducible. So, we proved the following
analogue of Theorem~\ref{thm3}.

\begin{Proposition}
\label{prop20}
Suppose that (\ref{10.1}) and (\ref{10.2}) hold
with monotone decreasing sequences $b= (b_n)$ and $  c=(c_n), $
and with $\alpha <0. $
Then the corresponding Spectral Riemann Surface is irreducible.
\end{Proposition}

We have to admit that
with all variety of pairs $(L,B)$ for which we have proved
the SRS's irreducibility, we know no nontrivial
(i.e., beside the case where some entries $b_k $ or $c_k $ vanish,
or diagonal entries are multiple) example of a pair $(L,B)$
with a reducible SRS.
\vspace{3mm}

6. From $\alpha < 0 $ we can go to another direction,
i.e., consider $\alpha \in (1/2, 1). $
The estimate (\ref{35.5}) is our conjecture, but even now we can claim the
following amendment to Theorem~\ref{thm1}.

\begin{Proposition}
\label{prop21}
Under the assumptions (\ref{10.1}) and (\ref{10.2}),
if $ \; 0 \leq  \alpha < 9/10, \; $
then the regularized trace
\begin{equation}
\label{}
tr_1 (z) = \sum_{n=1}^\infty \left ( E_n (z) - n^2 - \frac{1}{2}
E_n^{\prime \prime} (0)  z^2  \right )
\end{equation}
is well defined as an entire function of $z,$
and
\begin{equation}
tr_1 (z) \equiv 0.
\end{equation}
\end{Proposition}

The proof is based on (\ref{b36})--(\ref{b39}) and the estimates given by
Lemma \ref{lem31}. It
goes along the same lines as
Definition of regularized trace in Section 5.4 and
the
proof of Theorem~\ref{thm1};
see (\ref{60.1}) to (\ref{60.3}).
We omit the details.

Of course, one can introduce the higher order regularized traces
$$
tr_p (z) = \sum_{n=1}^\infty \left ( E_n (z) - n^2 - \sum_{j=1}^p
\frac{E_n^{(2j)} (0)}{(2j)!}  z^{2j}  \right )
$$
and study for which $\alpha $ this expression
is well defined as
an entire function.

It is important to mention that
many interesting examples of evaluation of
a regularized trace can be found in the recent
papers \cite{FM1,FM2,FM3,KP,MF,SP,Sav,SS1}
although there the operators $L$ and $B$ are usually self--adjoint
and $z $ is real. Let us notice that, in our 
Theorem \ref{thm1},
the first line of (\ref{7.2}), $ \alpha < 1/2,$
can be interpreted as an example to
Thm 1 in \cite{SP}.
Then, the second line of (\ref{7.2}) shows that
the restrictions on
$\delta$ and $\omega$ in \cite{SP}
could not be weakened.

\end{document}